\documentclass[12pt,a4paper,superscriptaddress]{revtex4}
\usepackage[latin1]{inputenc}
\usepackage[dvipsnames]{xcolor}
\usepackage{graphicx}
\usepackage{amsmath}
\usepackage{amssymb}
\usepackage{dcolumn}
\usepackage{bm}
\usepackage{hyperref}
\newcommand{\be}{\begin{equation}}
\newcommand{\ee}{\end{equation}}
\newcommand{\bea}{\begin{eqnarray}}
\newcommand{\eea}{\end{eqnarray}}

\DeclareRobustCommand{\revred}[1]{{#1}}
\DeclareRobustCommand{\revgreen}[1]{{#1}}

\begin{document}

\title{Four-dimensional supersymmetric massive QED}

\author{A. C. Lehum}
\email{lehum@ufpa.br}
\affiliation{Faculdade de F\'{i}sica, Universidade Federal do Par\'{a}\\ 66075-110, Bel\'{e}m, Par\'a, Brazil}

\author{J. R. Nascimento}
\email{jroberto@fisica.ufpb.br}
\affiliation{Departamento de F\'{\i}sica, Universidade Federal da 
Para\'{\i}ba\\
 Caixa Postal 5008, 58051-970, Jo\~ao Pessoa, Para\'{\i}ba, Brazil}

\author{A. Yu. Petrov}
\email{petrov@fisica.ufpb.br}
\affiliation{Departamento de F\'{\i}sica, Universidade Federal da 
Para\'{\i}ba\\
 Caixa Postal 5008, 58051-970, Jo\~ao Pessoa, Para\'{\i}ba, Brazil}

\begin{abstract}
We study a four-dimensional $\mathcal N=1$ supersymmetric Proca model with an explicit mass term for a real vector superfield and a generalized transverse kinetic operator. Although it displays no gauge redundancy in Proca variables, for local kinetic operators the model is the exact classical unitary-gauge form of the minimal Abelian $\mathcal N=1$ Stueckelberg realization, with the SQED-like exponential as its minimal gauge-covariant matter coupling. We compute the K\"ahlerian contribution to the one-loop effective action using both the background-field Hessian and an independent supergraph organization. In the minimal model the background-dependent divergence is local and is absorbed by the matter K\"ahler normalization, and the finite local terms left by the subtraction remain scheme dependent. At $\alpha=0$, an anomaly-free minimal Stueckelberg formulation supports a conditional perturbative-renormalizability interpretation for the corresponding minimal theory. For $f(\Box)=\Box+\alpha\Box^2$ we distinguish the exact polynomial model, defined over the full momentum range, from a truncated derivative expansion; in the latter reading, the additional pole and the local matching term are ultraviolet information determined by the completion rather than predictions of the truncated theory.
\end{abstract}
\maketitle

\section{Introduction}

{Four-dimensional supersymmetric gauge theories play a central role in quantum field theory. They have been studied in connection with the AdS/CFT correspondence, grand unified models, string theory, and the search for theories that are finite to all orders in perturbation theory \cite{Mald,Raby:2017ucc,Kovacs,Moha}. The paradigmatic example of the latter is ${\cal N}=4$ super-Yang--Mills theory \cite{Kovacs2}.}

{Supersymmetric realizations of a massive Abelian vector field, the supersymmetric analogue of the Maxwell--Proca theory, have been developed along several complementary lines in four dimensions. Ref.~\cite{Nishino:2013zxa} introduced a component $\mathcal{N}=1$ non-Abelian Proca--St\"uckelberg formulation in which a compensator multiplet gives mass to the vector multiplet while preserving supersymmetry. In a related $\mathcal{N}=1$ setting, one-loop corrections for complex linear superfields coupled to a massive Abelian vector generate higher-derivative superspace operators and deform the auxiliary-field potential~\cite{Farakos:2014iwa}.}

{In supergravity (SUGRA), massive vector multiplets also arise from nonlinear $D$-term dynamics and Dirac--Born--Infeld (DBI) structures. Ref.~\cite{Abe:2015fha} constructs an explicit four-dimensional $\mathcal{N}=1$ SUGRA massive-vector DBI action and discusses its dual relation to Starobinsky-type models. Ref.~\cite{Abe:2018plc} shows that a massive vector multiplet can consistently coexist with DBI kinetic terms and a new Fayet--Iliopoulos term without gauging $R$ symmetry, thereby placing the bosonic mass and nonlinear kinetic terms within a single multiplet.}

From a phenomenological perspective, following a Stueckelberg extension of the Standard Model~\cite{Kors:2004dx}, K\"ors and Nath constructed a supersymmetric Stueckelberg extension of the MSSM in which an additional Abelian vector multiplet and a chiral Stueckelberg multiplet combine into a massive vector multiplet~\cite{Kors:2004ri,Kors:2005uz}. In their model the Stueckelberg mass term also mixes the new vector multiplet with hypercharge, the MSSM fields are neutral under the new $U(1)_X$, and hidden-sector matter may carry $U(1)_X$ charge~\cite{Kors:2004ri,Kors:2005uz}. A recent analysis revisited the Abelian super Stueckelberg model in superspace and components, derived the supersymmetry transformations compatible with the Wess--Zumino gauge, and discussed extensions by higher powers of the gauge-invariant Stueckelberg superfield~\cite{Capri:2025SSM}.

The theory studied here is written in Proca variables: it is defined by an explicit mass term for a real vector superfield and, in these variables, displays no gauge redundancy. Its gauge-invariant parent is nevertheless a Stueckelberg construction. {For a local transverse kinetic operator, the Proca action with its minimal gauge-covariant matter coupling is the exact classical unitary-gauge form of the minimal Abelian $\mathcal N=1$ Stueckelberg realization, in which a chiral compensator restores supergauge invariance.} This relation is an exact change of variables, not a low-energy matching. A quantum-consistent gauge-invariant parent requires anomaly-free matter; this requirement and the relation to the K\"ors--Nath construction are discussed in Sec.~\ref{subsec:stueckelbergRealization}.

{We determine the K\"ahlerian part of the one-loop effective action of chiral background superfields, that is, the leading derivative-free full-superspace term, which fixes the one-loop correction to the matter K\"ahler metric; derivative-dependent and chiral contributions are not computed.} {In the minimal model the background-field Hessian and the supergraph expansion agree, and the background-dependent divergence is local and is absorbed by the matter K\"ahler normalization; for a general transverse form factor, what is established is the algebraic cancellation of the background-dependent longitudinal sector.} {At $\alpha=0$, the corresponding anomaly-free minimal Stueckelberg formulation supports a conditional perturbative-renormalizability interpretation, whose conditions are set out in Sec.~\ref{subsec:renormScheme}.} {For $f(\Box)=\Box+\alpha\Box^2$ we distinguish two uses of the same operator: an exact polynomial model over the full momentum range, which we analyze in closed form, and a truncated derivative expansion valid only at low momenta, whose predictions do not extend to the ultraviolet-sensitive features of the exact model.} {For this higher-derivative sector the effective-field-theory (EFT) question is therefore attached to the interpretation of the $\alpha\Box^2$ operator, not to the exact minimal Proca--Stueckelberg correspondence.}

{The supergraph ring expansion, projector algebra, and logarithmic resummation used here are standard tools in four-dimensional $\mathcal N=1$ calculations of effective K\"ahler potentials \cite{YMEP1,YMEP2,YMEP3,Brignole:2000,GrootNibbelink:2005,HDgauge}; we do not claim novelty for these methods. {The model-specific content is instead the one-loop K\"ahlerian determinant generated by an explicit vector-superfield Proca mass, which requires the complete real vector superfield to be retained.} {Ref.~\cite{HDgauge} considers a gauge-invariant higher-derivative Abelian theory in which gauge fixing determines the longitudinal sector, whereas here that sector is the unitary-gauge remnant of the Stueckelberg mass term and coexists with the transverse form factor; the difference is developed in Sec.~\ref{subsec:noveltyScope}.} The results below therefore provide a model-specific benchmark rather than a replacement for the general gauge-invariant literature on effective K\"ahler potentials.} For the general structure of effective actions in chiral superfield models, see, e.g., Ref.~\cite{BK0}. {We follow the superspace notation and supergraph conventions of Refs.~\cite{BK0,ourSUSY}, adapted throughout to the mostly-minus Minkowski metric $\eta_{ab}=\mathrm{diag}(+,-,-,-)$.}

{The paper is organized as follows. Section~\ref{sec:procaModel} introduces the supersymmetric Proca model, its free propagators, dynamical constraint, and coupling to chiral matter, together with its exact Stueckelberg realization. Section~\ref{sec:oneLoopKahler} presents the unified one-loop K\"ahlerian analysis, first for the minimal model and then for a general transverse form factor. {Section~\ref{sec:discussionScope} discusses renormalization and model scope, including the conditional Stueckelberg renormalizability interpretation and the distinction between the exact polynomial model and its truncated-EFT reading.} Section~\ref{sec:conclusions} summarizes the conclusions.}

\section{\revred{Supersymmetric Proca model}}
\label{sec:procaModel}

\subsection{\revred{Superspace conventions and projector algebra}}
\label{subsec:superspaceConventions}

We work in flat four-dimensional $\mathcal{N}=1$ Minkowski superspace, parametrized by
\begin{equation}
 z^M=(x^a,\theta^\alpha,\bar\theta^{\dot\alpha}),
 \qquad
 a=0,1,2,3,\quad \alpha,\dot\alpha=1,2,
\end{equation}
with metric $\eta_{ab}=\mathrm{diag}(+,-,-,-)$. We use the Fourier convention
\begin{equation}
 f(x)=\int\frac{d^4k}{(2\pi)^4}e^{-ik\cdot x}\widetilde f(k),
 \qquad
 \partial_a\longrightarrow-ik_a,
 \qquad
 \Box\longrightarrow-k^2,
\end{equation}
where $k^2=(k^0)^2-\mathbf{k}^2$. Consequently, $\Box+m^2\longrightarrow-(k^2-m^2)$, and the physical massive pole is located at $k^2=m^2$. The full and chiral superspace measures are denoted by
\begin{equation}
 d^8z\equiv d^4x\,d^2\theta\,d^2\bar\theta,
 \qquad
 d^6z\equiv d^4x\,d^2\theta,
\end{equation}
with the Grassmann integrations normalized by
$\int d^2\theta\,\theta^2=1$ and
$\int d^2\bar\theta\,\bar\theta^2=1$.
The flat superspace covariant derivatives are
\begin{equation}
 D_\alpha=\frac{\partial}{\partial\theta^\alpha}
 +i\bar\theta^{\dot\alpha}\partial_{\alpha\dot\alpha},
 \qquad
 \bar D_{\dot\alpha}=-\frac{\partial}{\partial\bar\theta^{\dot\alpha}}
 -i\theta^\alpha\partial_{\alpha\dot\alpha},
 \qquad
 \partial_{\alpha\dot\alpha}
 =\sigma^a_{\alpha\dot\alpha}\partial_a,
\end{equation}
and obey
\begin{equation}
 \{D_\alpha,\bar D_{\dot\alpha}\}
 =-2i\partial_{\alpha\dot\alpha},
 \qquad
 \{D_\alpha,D_\beta\}
 =\{\bar D_{\dot\alpha},\bar D_{\dot\beta}\}=0.
\end{equation}
We use $D^2\equiv D^\alpha D_\alpha$, $\bar D^2\equiv\bar D_{\dot\alpha}\bar D^{\dot\alpha}$, and $\Box\equiv\partial^a\partial_a$. Spinor indices are raised and lowered with the antisymmetric two-dimensional tensors.

For two superspace points, we define the purely Grassmannian delta function and the full superspace delta function by
\begin{equation}
 \delta_{12}\equiv
 \delta^2(\theta_1-\theta_2)\,
 \delta^2(\bar\theta_1-\bar\theta_2),
 \qquad
 \delta^8(z_1-z_2)
 =\delta^4(x_1-x_2)\,\delta_{12}.
\end{equation}
The notation
$\left.\delta_{12}\right|_{\theta_1=\theta_2}$ means that all spinor derivatives appearing in the corresponding kernel are first allowed to act on $\delta_{12}$ and only afterwards are $\theta_1=\theta_2$ and $\bar\theta_1=\bar\theta_2$ imposed. With the normalization above,
\begin{equation}
 \left.\frac{D^2\bar D^2}{16}\delta_{12}
 \right|_{\theta_1=\theta_2}=1.
\end{equation}
Functions of $\Box$ act on the bosonic factor $\delta^4(x_1-x_2)$ before the coincidence limit $x_1=x_2$ is taken.

The real vector superfield is denoted by $v=v^\dagger$, while $\phi$ and $\bar\phi$ denote a chiral matter superfield and its antichiral conjugate,
\begin{equation}
 \bar D_{\dot\alpha}\phi=0,
 \qquad
 D_\alpha\bar\phi=0.
\end{equation}
The parameters $g$ and $m$ denote, respectively, the dimensionless matter--vector coupling and the Proca mass. Their mass dimensions, together with those of the superfields used below, are
\begin{equation}
 [v]=0,\qquad [\phi]=[\bar\phi]=1,\qquad [g]=0,
 \qquad [m]=1.
\end{equation}
In the background-field expansion, $\Phi$ and $\bar\Phi$ denote the background chiral and antichiral superfields, whereas lowercase fields denote quantum fluctuations. The supergauge parameter $\Lambda$ is an arbitrary chiral superfield and $\bar\Lambda$ is its antichiral conjugate.

\subsection{\revred{Explicit Proca action and free propagators}}
\label{subsec:procaActionPropagators}

We begin by recalling the free Abelian SQED vector action. In the conventions of Refs.~\cite{BK0,ourSUSY}, it is
\bea
\label{action}
S_V=\frac{1}{4}\int
d^6zW^{\alpha}W_{\alpha}=-\frac{1}{16}\int d^8 z v
D^{\alpha} \bar{D}^2D_{\alpha}v,
\eea
where $v$ is a real vector superfield and $W_{\alpha}=\frac{1}{4}\bar{D}^2D_{\alpha}v$ is its chiral field-strength superfield. The action is invariant under $v\to v+\Lambda+\bar{\Lambda}$, with $\Lambda$ chiral and $\bar{\Lambda}$ antichiral. In the chiral representation of Eq.~\eqref{action}, the Hermitian-conjugate contribution is understood; the equivalent full-superspace form displayed there is manifestly real. In terms of the transverse superspace projector, the same action can be written as
\bea
S_V=\frac{1}{2}\int d^8z v\Pi_{1/2}\Box v,
\eea
{where $\Pi_{1/2}=-\frac{1}{8}\frac{D^{\alpha}\bar{D}^2D_{\alpha}}{\Box}$ is the transverse projector. The complementary longitudinal projector is $\Pi_0=\frac{1}{16}\frac{\{D^2,\bar{D}^2\}}{\Box}$. They satisfy}
$$
\Pi_{1/2}\Pi_{1/2}=\Pi_{1/2};\quad\, \Pi_0\Pi_0=\Pi_0;\quad\, \Pi_{1/2}\Pi_0=\Pi_0\Pi_{1/2}=0;\quad\, \Pi_{1/2}+\Pi_0=1.
$$ 
{At the component level, Eq.~\eqref{action} contains}
\bea
S_V=\int d^4x(-\frac{1}{4}F_{ab}F^{ab}+\ldots),
\eea
{where $F_{ab}=\partial_aA_b-\partial_bA_a$ is the Abelian }{field-strength tensor}{ of $A_a$, and the ellipsis denotes the fermionic and auxiliary-field terms of the vector multiplet.}

{We now add the explicit Proca mass term $S_m=\frac{m^2}{2}\int d^8z\,v^2$. Its vector component is $S_m=\frac{1}{2}\int d^4x\,m^2A_aA^a+\ldots$, and the complete superfield Proca action is therefore}
{
\begin{equation}
S_P=\frac{1}{2}\int d^8z\,
v\left[\Pi_{1/2}\Box+m^2\right]v
=\frac{1}{2}\int d^8z\,
v\left[\Pi_{1/2}(\Box+m^2)+\Pi_0m^2\right]v.
\end{equation}
}
{The corresponding vector-superfield propagator is}
{
\begin{align}
\label{propgauge}
\langle v(z_1)v(z_2)\rangle
&=-\left[
\frac{\Pi_{1/2}}{\Box+m^2}
+\frac{\Pi_0}{m^2}
\right]\delta^8(z_1-z_2)
\nonumber\\
&=-\frac{1}{\Box+m^2}
\left[
1+\frac{\{D^2,\bar D^2\}}{16m^2}
\right]\delta^8(z_1-z_2).
\end{align}
The second line follows from $\Pi_{1/2}+\Pi_0=1$ and makes the
analytic structure manifest. Although the definitions of the
superspace projectors contain $\Box^{-1}$, these factors do not generate
an additional pole at $\Box=0$: the only nonlocal denominator is
$\Box+m^2$. An equivalent decomposition is
\begin{equation}
\langle v(z_1)v(z_2)\rangle
=-\frac{1}{m^2}\delta^8(z_1-z_2)
-\frac{D^\alpha\bar D^2D_\alpha}
{8m^2(\Box+m^2)}\delta^8(z_1-z_2),
\end{equation}
which displays explicitly a local contact term and a propagating term
with denominator $\Box+m^2$. The singular massless limit arises not
from an additional massless pole, but from the explicit $m^{-2}$ factor
in the longitudinal sector.

\subsection{\texorpdfstring{On-shell constraint and spin-one propagator}{On-shell constraint and spin-one propagator}}
\label{subsec:procaConstraintStatus}

The spin-one component provides a direct check. Retaining the vector
field $A_a$ in the component action gives
\begin{equation}
S_A=\int d^4x\left(
-\frac14F_{ab}F^{ab}+\frac12m^2A_aA^a
\right)
=\frac12\int d^4x\,
A^a\left[(\Box+m^2)\eta_{ab}-\partial_a\partial_b\right]A^b.
\end{equation}
Introducing the transverse and longitudinal spacetime projectors,
\begin{equation}
P^T_{ab}=\eta_{ab}-\frac{\partial_a\partial_b}{\Box},
\qquad
P^L_{ab}=\frac{\partial_a\partial_b}{\Box},
\end{equation}
the inverse quadratic operator is
\begin{equation}
\mathcal O^{-1}_{ab}
=\frac{P^T_{ab}}{\Box+m^2}+\frac{P^L_{ab}}{m^2}
=\frac{1}{\Box+m^2}
\left(\eta_{ab}+\frac{\partial_a\partial_b}{m^2}\right).
\end{equation}
Thus, with $\partial_a\to-ik_a$, the Minkowski Feynman propagator is
\begin{equation}
D_{ab}(k)=\frac{-i}{k^2-m^2+i0}
\left(\eta_{ab}-\frac{k_ak_b}{m^2}\right),
\end{equation}
which is precisely the usual Proca propagator. It has no massless pole,
but its $m\to0$ limit is singular. Moreover, the longitudinal numerator
does not decrease as $1/k^2$ at large momentum, thereby reproducing the
worsened UV behavior characteristic of the Proca propagator.
These properties are encoded in the longitudinal superspace projector.
}

{The dynamical counterpart of the Lorenz condition can be stated directly in superspace. Varying the quadratic Proca action gives
\begin{equation}
\left(\Box\Pi_{1/2}+m^2\right)v=0.
\end{equation}
Since the transverse projector is annihilated by both $D^2$ and $\bar D^2$, applying these operators to the field equation yields, for $m^2\neq0$,
\begin{equation}
D^2v=0,
\qquad
\bar D^2v=0.
\label{linearconstraint}
\end{equation}
Thus, the massive real superfield satisfies the real-linear constraint on shell. Equivalently, $\Pi_0v=0$ and $\Pi_{1/2}v=v$, and the remaining field equation is $(\Box+m^2)v=0$. Its vector component, $A_{\alpha\dot\alpha}\propto[D_\alpha,\bar D_{\dot\alpha}]v|$, consequently satisfies
\begin{equation}
\partial^aA_a=0.
\end{equation}
This is the supersymmetric form of the Proca transversality condition. It follows dynamically from the massive field equation and is not a gauge choice.}

{The same conclusion holds in the presence of the chiral-matter coupling considered below. Defining
\begin{equation}
J=\bar\phi e^{gv}\phi,
\end{equation}
the vector-superfield equation becomes
\begin{equation}
\left(\Box\Pi_{1/2}+m^2\right)v+gJ=0.
\end{equation}
The matter equations,
\begin{equation}
-\frac14\bar D^2\left(\bar\phi e^{gv}\right)=0,
\qquad
-\frac14D^2\left(e^{gv}\phi\right)=0,
\end{equation}
imply
\begin{equation}
D^2J=\bar D^2J=0.
\end{equation}
Consequently, Eq.~\eqref{linearconstraint} follows from the complete classical equations of motion. In components, this result is equivalent to taking the divergence of the sourced Proca equation and using conservation of the matter current.}

{Although the Proca term explicitly breaks the Abelian supergauge invariance of the complete action, {its contribution to the Abelian Ward identity is linear in the vector superfield}, whereas the matter action is classically invariant under $v\to v+\Lambda+\bar\Lambda$, $\phi\to e^{-g\Lambda}\phi$. At one loop, the matter contribution to the vector two-point function is therefore transverse: for a single chiral superfield the Abelian anomaly functional, proportional to $\mathrm{Tr}\,q^3$, is quadratic in $v$, so its first functional derivative vanishes at $v=0$ and it first enters the three-point Ward identity. Here $\mathrm{Tr}\,q^n\equiv\sum_i q_i^n$ denotes the trace over the left-handed Weyl fermions in the charged chiral multiplets, with $q_i$ their Abelian charges in the normalization of Eq.~\eqref{stueckelbergAction}.  A single chiral superfield with $q=g$ therefore gives $\mathrm{Tr}\,q^3=g^3$ and $\mathrm{Tr}\,q=g$, corresponding respectively to the cubic gauge and mixed gauge--gravitational anomaly coefficients, whereas the Stueckelberg fermion is neutral and does not contribute~\cite{Kors:2004ri,Kors:2005uz}.
In particular, the quadratic vector-superfield 1PI operator has the form
\begin{equation}
\Gamma^{(2)}_{vv}
=
\Pi_{1/2}\left[\Box+m_R^2+\Sigma_T(\Box)\right]+m_R^2\Pi_0,
\end{equation}
where $\Sigma_T(\Box)$ is the transverse vacuum-polarization form factor. In this one-loop two-point sector, no independent radiative longitudinal form factor is generated. Applying $D^2$ and $\bar D^2$ to the quantum field equation therefore gives
\begin{equation}
D^2v=\bar D^2v=0
\end{equation}
for nonzero renormalized Proca mass. Thus, in the one-loop two-point sector, matter radiative corrections modify the transverse dynamics but preserve the dynamical Lorenz condition. Any all-order statement would require an anomaly-free formulation, such as the embedding of Sec.~\ref{subsec:stueckelbergRealization}, together with a separate analysis of the corresponding quantum Ward identity, which is outside the present calculation.}

{For completeness, the positivity of the minimal free spin-one sector can be displayed after eliminating the nondynamical component $A_0$. The resulting Hamiltonian density is
\begin{equation}
\mathcal H_A
=
\frac12\left(\mathbf E^2+\mathbf B^2\right)
+\frac12m^2\mathbf A^2
+\frac{1}{2m^2}\left(\boldsymbol\nabla\cdot\mathbf E\right)^2,
\end{equation}
which is nonnegative for $m^2>0$. At the superfield level, the only propagating pole lies in the $\Pi_{1/2}$ sector and has the conventional sign, whereas the $\Pi_0/m^2$ contribution is nonpropagating. The minimal quadratic theory therefore has the standard ghost-free massive-vector-multiplet spectrum. Supersymmetry organizes these physical states into a massive supermultiplet, but it does not replace the spectral positivity test. The longitudinal constraint persists in the one-loop two-point sector; a complete loop-level analysis of transverse spectral positivity would require the quantum vector two-point function and is not part of the K\"ahlerian calculation performed here.}

{We couple the chiral matter multiplet to the real Proca superfield through the manifestly supersymmetric interaction}
\begin{equation}
{S_c=\int d^8z\,\bar\phi e^{gv}\phi.}
\end{equation}
{This interaction has the familiar SQED form; no supergauge symmetry is manifest in the present Proca variables, but it is the minimal gauge-covariant coupling of the Stueckelberg realization discussed in Sec.~\ref{subsec:stueckelbergRealization}. Here $g$ is the matter--vector coupling, $\phi$ is chiral, $\bar D_{\dot\alpha}\phi=0$, and $\bar\phi$ is antichiral, $D_\alpha\bar\phi=0$.}
{The free chiral-superfield propagators are}
\bea
\label{props}
&&<\phi(z_1)\bar{\phi}(z_2)>=\frac{\bar{D}^2D^2}{16\Box}\delta^8(z_1-z_2);\,
<\bar{\phi}(z_1)\phi(z_2)>=\frac{D^2\bar{D}^2}{16\Box}\delta^8(z_1-z_2);
\eea
{Their sum satisfies $<\phi(z_1)\bar{\phi}(z_2)>+<\bar{\phi}(z_1)\phi(z_2)>=\Pi_0\delta^8(z_1-z_2)$, an identity used below in the mixed-ring calculation.}

\subsection{\texorpdfstring{Stueckelberg realization and scope of the Proca action}{Stueckelberg realization and scope of the Proca action}}
\label{subsec:stueckelbergRealization}

The explicit Proca action of Sec.~\ref{subsec:procaActionPropagators} admits an exact gauge-invariant reformulation. Consider a local transverse kinetic function $f(\Box)=\Box\,h(\Box)$, generated by $\int d^6z\,W^\alpha h(\Box)W_\alpha+\mathrm{h.c.}$ with the normalization of Eq.~\eqref{action}, and a dimensionless chiral compensator $\Sigma$, $\bar D_{\dot\alpha}\Sigma=0$. 
The action
\begin{equation}
S_{\rm St}=\frac12\int d^8z\,v\,f(\Box)\Pi_{1/2}v+\frac{m^2}{2}\int d^8z\,(v-\Sigma-\bar\Sigma)^2+\int d^8z\,\bar\phi e^{gv}\phi
\label{stueckelbergAction}
\end{equation}
is invariant under $v\to v+\Lambda+\bar\Lambda$, $\Sigma\to\Sigma+\Lambda$, $\phi\to e^{-g\Lambda}\phi$, because $\Pi_{1/2}(\Lambda+\bar\Lambda)=0$. 
Since $\int d^8z\,\Sigma^2=0$, the mass term contains the compensator kinetic term $m^2\int d^8z\,\bar\Sigma\Sigma$. 
In terms of the gauge-invariant superfields $V=v-\Sigma-\bar\Sigma$ and $\phi'=e^{g\Sigma}\phi$, one finds identically
\begin{equation}
S_{\rm St}=\frac12\int d^8z\,V\bigl[f(\Box)\Pi_{1/2}+m^2\bigr]V+\int d^8z\,\bar\phi' e^{gV}\phi' ,
\label{stueckelbergToProca}
\end{equation}
which is the Proca action with the matter coupling used in this paper. 
Equivalently, the unitary gauge $\Lambda=-\Sigma$, reached algebraically and without residual transformations, sets $\Sigma=0$. 
This is an exact classical change of variables; it is not a low-energy approximation, and no matching is involved. 
For $f(\Box)=\Box$, Eq.~\eqref{stueckelbergAction} reproduces, after translation of conventions, the Abelian super Stueckelberg action of Ref.~\cite{Capri:2025SSM} with $S=i\Sigma$, for which their invariant superfield $V^S$ coincides with $V$. 
It likewise reproduces the $C$--$S$ sector of the $M_2=0$ truncation of the K\"ors--Nath construction, with $S=-M_1\Sigma$ and $M_1\leftrightarrow m$, when their hidden-sector matter is directly charged under $U(1)_X$ and $g=2g_XQ_X$~\cite{Kors:2004ri,Kors:2005uz}. 

Because $\Pi_0(\Sigma+\bar\Sigma)=\Sigma+\bar\Sigma$, one has $\Pi_0V=\Pi_0v-\Sigma-\bar\Sigma$.
The algebraic block $m^2\Pi_0$ is therefore the longitudinal remnant of the Stueckelberg mass term in unitary gauge; it is neither the compensator nor its determinant, and the Faddeev--Popov determinant of this gauge is field independent.
Unitary gauge also differs from the Wess--Zumino gauge, which removes the lowest components of the vector superfield and leaves a residual ordinary $U(1)$ invariance that still requires gauge fixing~\cite{Capri:2025SSM}; unitary gauge uses the full chiral parameter, retains all components of $V$, and leaves no residual freedom. 

The full K\"ors--Nath model contains two Abelian factors, Stueckelberg mass mixing with hypercharge, and MSSM fields that are neutral under the new $U(1)$~\cite{Kors:2004ri,Kors:2005uz}; it is not identified with the present model. 
Quantum consistency of the gauge-invariant realization requires anomaly cancellation. 
A single chiral superfield of charge $g$ gives $\mathrm{Tr}\,q^3=g^3$ and $\mathrm{Tr}\,q=g$, whereas the Stueckelberg fermion is neutral and does not contribute~\cite{Kors:2004ri,Kors:2005uz}. 
An illustrative anomaly-free embedding adds a spectator $\phi_-$ of charge $-g$ with vanishing background and no superpotential coupling to $\phi$. 
Its fluctuations enter the quadratic action only through $\int d^8z\,\bar\varphi_-\varphi_-$, so the background-dependent one-loop Hessian of Sec.~\ref{subsec:backgroundHessian}, and with it $K^{(1)}(\Phi)$, is unchanged. 

The exponential is the minimal gauge-covariant coupling of charged chiral matter; what is not unique is the possible addition of non-minimal invariant interactions, such as higher powers of $V$~\cite{Capri:2025SSM} or further K\"ahler couplings between $V$ and $\bar\phi'\phi'$. 
The results of Sec.~\ref{sec:oneLoopKahler} refer to the minimal realization of Eq.~\eqref{stueckelbergToProca}. 
When a completion contains additional heavy fields that are integrated out, the resulting local vector-sector operators have Wilson coefficients that must be fixed by matching to that completion.

The higher-derivative part of $f(\Box)=\Box+\alpha\Box^2$ corresponds to the gauge-invariant operator $\int d^6z\,W^\alpha\Box W_\alpha+\mathrm{h.c.}$, with coefficient proportional to $\alpha$, and leaves Eq.~\eqref{stueckelbergToProca} intact. 
For the higher-derivative sector considered here, the EFT question is therefore attached to the interpretation of the $\alpha\Box^2$ operator, not to the exact minimal Proca--Stueckelberg correspondence.
The closed result of Sec.~\ref{subsec:generalFormFactor} is the exact one-loop K\"ahlerian result of the chosen polynomial model, integrated over the full Euclidean range.
If $\alpha\Box^2$ is instead regarded as the leading term of a derivative expansion valid for $|\alpha k^2|\ll1$, the apparent additional pole and the region $k_E^2\sim1/|\alpha|$ lie outside the controlled domain of the truncated theory and depend on its UV completion.
The local term in Eq.~\eqref{hdMinimalMatchingLimit} is then UV/threshold matching information rather than a prediction of the truncated theory. 

\section{\revred{One-loop K\"ahlerian effective action}}
\label{sec:oneLoopKahler}

\subsection{\revred{Scope of the K\"ahlerian approximation}}
\label{subsec:kahlerScope}

{
The one-particle-irreducible effective action
$\Gamma[\Phi,\bar\Phi]$ is a functional of the background chiral
superfields and, in general, contains more information than an
effective potential. In a superspace derivative expansion we write
schematically
\begin{equation}
\Gamma[\Phi,\bar\Phi]
=
\Gamma_K[\Phi,\bar\Phi]
+\Gamma_{\mathrm{der}}[\Phi,\bar\Phi]
+\left(\Gamma_{\mathrm{ch}}[\Phi]+\mathrm{h.c.}\right),
\end{equation}
Here $\Gamma_{\mathrm{der}}[\Phi,\bar\Phi]$ denotes the collection of full-superspace contributions containing spacetime or supercovariant derivatives acting on the background superfields, including derivative-dependent local terms and, more generally, nonlocal form factors. The functional $\Gamma_{\mathrm{ch}}[\Phi]$ denotes the chiral sector of the effective action, which can be written schematically as $\Gamma_{\mathrm{ch}}[\Phi]=\int d^6z\,\mathcal{L}_{\mathrm{ch,eff}}(\Phi,D\Phi,\partial\Phi,\ldots)$; its Hermitian conjugate is the corresponding antichiral contribution. This sector includes, when generated, effective-superpotential and chiral higher-derivative terms. Neither $\Gamma_{\mathrm{der}}$ nor $\Gamma_{\mathrm{ch}}$ is evaluated in the present work. The K\"ahlerian sector is defined by
\begin{equation}
\Gamma_K[\Phi,\bar\Phi]
=
\int d^8z\,K_{\mathrm{eff}}(\Phi,\bar\Phi)
\end{equation}
and consists of the local full-superspace
contribution containing no spacetime or supercovariant derivatives
acting on the background superfields. In the present work we compute
only the one-loop contribution $K^{(1)}$ to this sector,
$\Gamma_K^{(1)}=\int d^8z\,K^{(1)}$, and do not evaluate derivative,
nonlocal, or chiral contributions to the full effective action. The
backgrounds satisfy $\bar D_{\dot\alpha}\Phi=0$ and
$D_\alpha\bar\Phi=0$ and are treated in the K\"ahlerian approximation:
terms with derivatives acting on them are consistently omitted.}

\subsection{\revred{Background-field expansion and quadratic Hessian}}
\label{subsec:backgroundHessian}

{To make the background-field expansion and the one-loop combinatorics explicit, we distinguish the full matter superfields from their quantum fluctuations by writing}
\begin{equation}
{
\phi_{\rm tot}=\Phi+\varphi,
\qquad
\bar\phi_{\rm tot}=\bar\Phi+\bar\varphi,
\qquad
v_{\rm tot}=v,
}
\label{backgroundsplit}
\end{equation}
{where the vector background is set to zero because the effective action considered here depends only on the chiral backgrounds. Expanding the matter interaction gives}
\begin{equation}
{S_c=S_c^{(0)}+S_c^{(1)}+S_c^{(2)}+O({\cal Q}^3),}
\end{equation}
{where ${\cal Q}$ denotes collectively the quantum fields and}
\begin{align}
{S_c^{(0)}}&{=\int d^8z\,\bar\Phi\Phi,}\\
{S_c^{(1)}}&{=\int d^8z\left(
\bar\Phi\varphi+\bar\varphi\Phi+g\bar\Phi\Phi\,v\right),}\\
{S_c^{(2)}}&{=\int d^8z\left[
\bar\varphi\varphi+g\bar\Phi\,v\varphi+g\bar\varphi\,v\Phi
+\frac{g^2}{2}\bar\Phi\Phi\,v^2\right].}
\label{quadraticmatter}
\end{align}
{The linear term is cancelled by the sources entering the Legendre transform that defines the 1PI effective action; the backgrounds therefore need not satisfy the classical equations of motion. Terms cubic and higher in the quantum fields do not belong to the Hessian and start contributing beyond one loop. Consequently, the complete quadratic action relevant to the present calculation is}
\begin{align}
{S^{(2)}}={}&{\frac12\int d^8z\,
 v\left(\mathcal K_v+M^2\right)v
+\int d^8z\,\bar\varphi\varphi}\nonumber\\
&{+g\int d^8z\left(
\bar\Phi\,v\varphi+\bar\varphi\,v\Phi\right),}
\label{quadraticaction}
\end{align}
{where}
\begin{equation}
{
\mathcal K_v=\Box\Pi_{1/2}+m^2,
\qquad
M^2\equiv g^2\Phi\bar\Phi.
}
\end{equation}
{Within the K\"ahlerian approximation, derivatives acting on $\Phi$ and $\bar\Phi$ are omitted, so that $M^2$ can be commuted through the projector kernels and functions of $\Box$.}

{Equation~\eqref{quadraticaction} fixes the background-dependent vertices unambiguously. The two-vector insertion is $V_{vv}=M^2=g^2\Phi\bar\Phi$: the factor $1/2$ in $\frac12M^2v^2$ is cancelled by the two functional derivatives with respect to the identical vector-superfield fluctuations. The mixing vertices are $V_{v\varphi}=g\bar\Phi$ and $V_{v\bar\varphi}=g\Phi$. They are conventionally referred to as cubic and quartic vertices when the external background legs are counted, although all of them belong to the quadratic fluctuation operator. The genuinely cubic quantum vertex $g\bar\varphi v\varphi$ does not enter the one-loop determinant.}

\subsection{\revred{Common determinant and supergraph organization}}
\label{subsec:commonDeterminant}

{The real-field Gaussian determinant supplies a factor $1/2$, whereas cyclic invariance of a ring with $n$ repeated insertions supplies the factor $1/n$. The pure-vector rings therefore carry the symmetry factor $1/(2n)$. A mixed ring with $n$ repeated links has the same factor; the two chiral orientations are represented by $G_{\varphi\bar\varphi}+G_{\bar\varphi\varphi}=\Pi_0\delta^8$. Equivalently, the two diagram classes follow from the block determinant of the quadratic Hessian: the vector block generates the pure-vector rings, while its Schur complement with the chiral block generates the mixed rings.}

\begin{figure}[htbp] 
	\begin{center} 
		\includegraphics[width={10cm}]{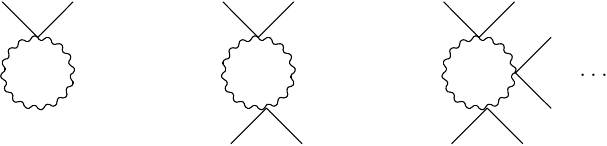} 
	\end{center} 
	\caption{Pure vector-superfield ring diagrams.}
	\label{Fig:diagrams1} 
\end{figure}

{We now evaluate the one-loop K\"ahlerian contribution.} {The block structure of Eq.~\eqref{quadraticaction} separates the one-loop determinant into two diagram classes, in agreement with the standard supergraph organization of Refs.~\cite{YMEP1,YMEP2,YMEP3,HDgauge}. The first class contains only vector-superfield propagators and the background-dependent two-vector insertion; its diagrams are displayed in Fig.~\ref{Fig:diagrams1}.}

\vspace*{2mm}

\vspace*{2mm}

\begin{figure}[htbp]
  \begin{center}
    \includegraphics[width={10cm}]{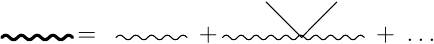}
  \end{center}
  \caption{Background-dressed vector-superfield propagator.}
  \label{Fig:diagrams2}
\end{figure}

\begin{figure}[htbp] 
	\begin{center} 
		\includegraphics[width={5cm}]{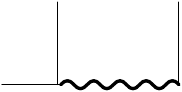} 
	\end{center} 
	\caption{Elementary link in a mixed ring.}
	\label{Fig:diagrams3} 
\end{figure}

\begin{figure}[htbp] 
	\begin{center} 
		\includegraphics[width={10cm}]{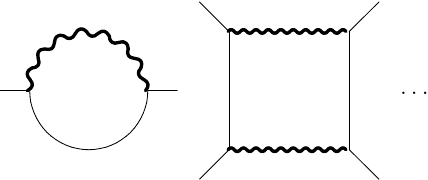} 
	\end{center} 
	\caption{Closed mixed-ring diagrams.}
	\label{Fig:diagrams4} 
\end{figure}

{Figures~\ref{Fig:diagrams1}--\ref{Fig:diagrams4} collect the supergraph topologies common to both kinetic realizations: pure-vector rings, the dressed vector propagator, an elementary mixed link, and the corresponding closed mixed rings. They are introduced here once and are not repeated in the general-$f(\Box)$ subsection.}

\vspace*{2mm}

\vspace*{2mm}

\vspace*{2mm}

\vspace*{2mm}

\subsection{\revred{Minimal Proca kinetic operator}}
\label{subsec:minimalProcaLoop}

\subsubsection{\revred{Pure-vector determinant}}
\label{subsubsec:minimalPureVector}

{The pure-vector contribution follows directly from the background-dependent part of the vector-superfield determinant. Subtracting the field-independent determinant at $M^2=0$, we have}
\begin{equation}
{
\Gamma^{(1)}_{K,a}
=\frac{i}{2}\operatorname{Tr}
\ln\left[(\mathcal K_v+M^2)\mathcal K_v^{-1}\right]
=\frac{i}{2}\operatorname{Tr}
\ln\left(1+M^2\mathcal K_v^{-1}\right),
}
\label{vectorDeterminant}
\end{equation}
{where the trace includes integration over the full superspace. Expanding the logarithm gives}
\begin{align}
{
\Gamma^{(1)}_{K,a}}
={}&{
\frac{i}{2}\int d^8z_1
\sum_{n=1}^{\infty}\frac{(-1)^{n+1}}{n}(M^2)^n}
\nonumber\\
&{\times
\left[
\frac{\Pi_{1/2}}{(\Box+m^2)^n}
+\frac{\Pi_0}{m^{2n}}
\right]
\delta^8(z_1-z_2)\Big|_{z_2=z_1}.}
\label{vectorRingSeries}
\end{align}
{All differential operators in Eq.~\eqref{vectorRingSeries} act to the right on the full delta distribution before the coincidence limit is taken. In the K\"ahlerian approximation, $M^2=g^2\Phi\bar\Phi$ is a multiplicative background and commutes with the projectors and functions of $\Box$. Writing $\delta^8(z_1-z_2)=\delta^4(x_1-x_2)\delta_{12}$, the spinor derivatives saturate the Grassmannian delta, while $\Box$ and its inverse act on the bosonic delta. In particular,}
\begin{equation}
{
\left.\Pi_0\delta^8(z_1-z_2)\right|_{\theta_1=\theta_2}
=\frac{2}{\Box}\delta^4(x_1-x_2),
\qquad
\left.\Pi_{1/2}\delta^8(z_1-z_2)\right|_{\theta_1=\theta_2}
=-\frac{2}{\Box}\delta^4(x_1-x_2).
}
\label{projectorCoincidence}
\end{equation}
{Consequently, Eq.~\eqref{vectorRingSeries} reduces to}
\begin{align}
{
\Gamma^{(1)}_{K,a}}
={}&{
i\int d^8z_1\sum_{n=1}^{\infty}
\frac{(-1)^n}{n}\frac{1}{\Box}
\left[
\left(\frac{M^2}{\Box+m^2}\right)^n
-\left(\frac{M^2}{m^2}\right)^n
\right]}
\nonumber\\
&{\times
\delta^4(x_1-x_2)\Big|_{x_2=x_1}.}
\label{vectorRingDA}
\end{align}
{Using $\sum_{n=1}^{\infty}(-1)^n x^n/n=-\ln(1+x)$, the resummed contribution is}
\begin{align}
{
\Gamma^{(1)}_{K,a}}
={}&{
-i\int d^8z_1\frac{1}{\Box}
\left[
\ln\left(1+\frac{M^2}{\Box+m^2}\right)
-\ln\left(1+\frac{M^2}{m^2}\right)
\right]}
\nonumber\\
&{\times
\delta^4(x_1-x_2)\Big|_{x_2=x_1}.}
\label{vectorRingLog}
\end{align}
{No infinite sum remains after this resummation. Since Eqs.~\eqref{vectorDeterminant}--\eqref{vectorRingLog} already contain the full superspace integral, they define a contribution to the effective action, denoted by $\Gamma^{(1)}_{K,a}$. The corresponding local K\"ahlerian density $K_a^{(1)}$ is defined by $\Gamma^{(1)}_{K,a}=\int d^8z\,K_a^{(1)}$.}

{We next evaluate the bosonic momentum trace. It is useful to define}
\begin{equation}
{
\Omega^2\equiv m^2+M^2=m^2+g^2\Phi\bar\Phi .
}
\label{OmegaDef}
\end{equation}
{With the Fourier and mostly-minus conventions stated in Sec.~2, Eq.~\eqref{vectorRingLog} becomes}
\begin{align}
{
\Gamma^{(1)}_{K,a}}
={}&{
i\int d^8z\int\frac{d^4k}{(2\pi)^4}\frac{1}{k^2+i0}
\left[
\ln\frac{k^2-\Omega^2+i0}{k^2-m^2+i0}
-\ln\frac{\Omega^2}{m^2}
\right].}
\label{vectorMinkowskiMomentum}
\end{align}
{Under the Wick rotation $k^0=ik_E^0$, one has $d^4k=i\,d^4k_E$ and $k^2=-k_E^2$. In particular,}
\begin{equation}
{
\int\frac{d^4k}{(2\pi)^4}\frac{F(k^2)}{k^2+i0}
=-i\int\frac{d^4k_E}{(2\pi)^4}\frac{F(-k_E^2)}{k_E^2}.
}
\label{WickIdentity}
\end{equation}
{The factor $-i$ in Eq.~\eqref{WickIdentity} cancels the explicit $+i$ in Eq.~\eqref{vectorMinkowskiMomentum}. Thus the K\"ahlerian density is real and carries no residual factor of $i$:}
\begin{equation}
{
K_a^{(1)}
=\mu^{\epsilon}\int\frac{d^d k_E}{(2\pi)^d}\frac{1}{k_E^2}
\left[
\ln\frac{k_E^2+\Omega^2}{k_E^2+m^2}
-\ln\frac{\Omega^2}{m^2}
\right],
\qquad d=4-\epsilon .
}
\label{res0}
\end{equation}
{We use dimensional reduction: the superspace and spinor-derivative algebra remain four-dimensional, whereas only the loop momentum is continued to $d=4-\epsilon$. We define}
\begin{equation}
{
\Delta_{\overline{\mathrm{DR}}}
\equiv\frac{2}{\epsilon}-\gamma_E+\ln(4\pi).
}
\label{DREDDelta}
\end{equation}
{The second logarithm in Eq.~\eqref{res0} multiplies the scaleless integral $\mu^{\epsilon}\int d^d k_E/[(2\pi)^d k_E^2]$ and therefore vanishes in DRED. The remaining integral can be evaluated by differentiating with respect to $\Omega^2$:}
\begin{align}
{
\frac{\partial K_a^{(1)}}{\partial\Omega^2}}
={}&{
\mu^{\epsilon}\int\frac{d^d k_E}{(2\pi)^d}
\frac{1}{k_E^2(k_E^2+\Omega^2)} }
\nonumber\\
={}&{
\frac{\mu^{\epsilon}}{(4\pi)^{2-\epsilon/2}}
\frac{\Gamma(\epsilon/2)}{1-\epsilon/2}
(\Omega^2)^{-\epsilon/2}.}
\label{vectorIntegralDerivative}
\end{align}
{Imposing the determinant normalization $K_a^{(1)}(\Omega^2=m^2)=0$ and expanding around $\epsilon=0$, one obtains}
\begin{align}
{
K_a^{(1)}}
={}&{
\frac{1}{16\pi^2}
\bigg\{
\Omega^2\left[
\Delta_{\overline{\mathrm{DR}}}+2
-\ln\frac{\Omega^2}{\mu^2}
\right]
-m^2\left[
\Delta_{\overline{\mathrm{DR}}}+2
-\ln\frac{m^2}{\mu^2}
\right]
\bigg\}.}
\label{res}
\end{align}
{The $m^2$ term is independent of the background superfields and may be omitted. The field-dependent pole is}
\begin{equation}
{
K_{a,\mathrm{div}}^{(1)}
=\frac{g^2\Phi\bar\Phi}{16\pi^2}
\Delta_{\overline{\mathrm{DR}}},
}
\label{vectorDivergence}
\end{equation}
{and is removed at this order by the local K\"ahler counterterm $\delta K_a=-g^2\Phi\bar\Phi\,\Delta_{\overline{\mathrm{DR}}}/(16\pi^2)$. Minimal $\overline{\mathrm{DR}}$ subtraction then gives}
\begin{align}
{
K_{a,\overline{\mathrm{DR}}}^{(1)}}
={}&{
\frac{1}{16\pi^2}
\left\{
\Omega^2\left[2-\ln\frac{\Omega^2}{\mu^2}\right]
-m^2\left[2-\ln\frac{m^2}{\mu^2}\right]
\right\}.}
\label{res1}
\end{align}
{A further finite local wave-function counterterm can change the term proportional to $g^2\Phi\bar\Phi$, but not the sign of the nonpolynomial logarithmic dependence. The result above renormalizes the one-loop K\"ahlerian sector. {The renormalizability status of the complete theory is discussed in Sec.~\ref{subsec:renormScheme}.}}

\subsubsection{\revred{Dressed vector propagator and mixed determinant}}
\label{subsubsec:minimalMixed}

{
The mixed vector--chiral contribution is most conveniently organized by
first resumming the background-dependent two-vector insertions along an
open vector-superfield line. This resummation is represented schematically
in Fig.~\ref{Fig:diagrams2}.
}

{
From Eq.~\eqref{quadraticaction}, the relevant vector operator is
\begin{align}
\mathcal K_v+M^2
&=\Box\Pi_{1/2}+m^2(\Pi_{1/2}+\Pi_0)
 +M^2(\Pi_{1/2}+\Pi_0)\nonumber\\
&=(\Box+\Omega^2)\Pi_{1/2}+\Omega^2\Pi_0,
\label{dressedVectorOperator}
\end{align}
where $\Omega^2=m^2+M^2$. Using the orthogonality and idempotence of
$\Pi_{1/2}$ and $\Pi_0$, its inverse is
\begin{equation}
(\mathcal K_v+M^2)^{-1}
=\frac{\Pi_{1/2}}{\Box+\Omega^2}
 +\frac{\Pi_0}{\Omega^2}.
\label{dressedVectorInverse}
\end{equation}
The overall sign is fixed by the free propagator in
Eq.~\eqref{propgauge}. Therefore, the background-dressed two-point
kernel is
\begin{align}
\langle v(z_1)v(z_2)\rangle_D
&=-\left[
\frac{\Pi_{1/2}}{\Box+\Omega^2}
+\frac{\Pi_0}{\Omega^2}
\right]\delta^8(z_1-z_2)
\label{dressedVectorPropagator}\\
&=-\frac{1}{\Box+\Omega^2}
\left[1+\frac{\{D^2,\bar D^2\}}{16\Omega^2}\right]
\delta^8(z_1-z_2).
\nonumber
\end{align}
The second form shows that the inverse powers of $\Box$ contained in
the projectors do not introduce an additional propagating pole at
$\Box=0$. The only transverse pole is determined by
$\Box+\Omega^2=0$, or $k^2=\Omega^2$ with the Fourier conventions
used here, whereas the longitudinal sector has the algebraic denominator
$\Omega^2$ and is nonpropagating.

As an independent check, let
\begin{equation}
\mathcal A\equiv\mathcal K_v^{-1}
=\frac{\Pi_{1/2}}{\Box+m^2}+\frac{\Pi_0}{m^2},
\qquad
G_0=-\mathcal A.
\end{equation}
Within the K\"ahlerian approximation, $M^2$ commutes with the
projector kernels. The open-chain Dyson series is then
\begin{align}
G_D
&=G_0+G_0M^2G_0+G_0M^2G_0M^2G_0+\cdots\nonumber\\
&=G_0\sum_{n=0}^{\infty}(M^2G_0)^n
=-\sum_{n=0}^{\infty}(-1)^n(M^2)^n\mathcal A^{n+1}.
\label{dressedVectorGeometricSeries}
\end{align}
All products in Eq.~\eqref{dressedVectorGeometricSeries} denote
convolutions over intermediate superspace points, and all differential
operators act on the full superspace delta distribution. Projector
orthogonality gives
\begin{equation}
\mathcal A^{n+1}
=\frac{\Pi_{1/2}}{(\Box+m^2)^{n+1}}
 +\frac{\Pi_0}{m^{2(n+1)}},
\end{equation}
so that the two geometric sums reproduce
Eq.~\eqref{dressedVectorPropagator} separately in the transverse and
longitudinal sectors.

In particular, the longitudinal projection is
\begin{equation}
\boxed{
\Pi_0\langle v(z_1)v(z_2)\rangle_D
=-\frac{\Pi_0}{\Omega^2}\delta^8(z_1-z_2)
=-\frac{\Pi_0}{m^2+M^2}\delta^8(z_1-z_2).
}
\label{dressedLongitudinalProjection}
\end{equation}
The limit $M^2\to0$ reproduces Eq.~\eqref{propgauge}. The result above
is an open two-point kernel and carries neither a cyclic factor
$1/n$ nor the real-field determinant factor $1/2$. Those factors
belong to the closed mixed-ring expansion, which must be derived
separately.
}

\vspace*{2mm}

{The closed mixed rings shown in Fig.~\ref{Fig:diagrams4} can be derived directly from the Schur complement of the quadratic Hessian and independently reproduced by a supergraph expansion. Define the two oriented chiral projectors}
\begin{equation}
{
P_+\equiv\frac{\bar D^2D^2}{16\Box},
\qquad
P_-\equiv\frac{D^2\bar D^2}{16\Box},
\qquad
P_++P_-=\Pi_0,
}
\label{mixedChiralProjectors}
\end{equation}
{where
$G_{\varphi\bar\varphi}=P_+\delta^8$ and
$G_{\bar\varphi\varphi}=P_-\delta^8$. The two terms represent the two globally consistent chiral orientations. Since $P_+P_-=P_-P_+=0$ and $P_\pm^2=P_\pm$, their sum includes both orientations without generating an additional factor of two. In the K\"ahlerian approximation, the two positive mixing vertices produce the vector-to-vector matter bridge}
\begin{align}
{
\Sigma_\chi(z_1,z_2)}
={}&{
g^2\bar\Phi(z_1)G_{\varphi\bar\varphi}(z_1,z_2)\Phi(z_2)
+g^2\Phi(z_1)G_{\bar\varphi\varphi}(z_1,z_2)\bar\Phi(z_2)}
\nonumber\\
\xrightarrow[\text{K\"ahlerian}]{}&
{M^2\Pi_0\delta^8(z_1-z_2).}
\label{mixedMatterBridge}
\end{align}

{Let
$\mathcal A=\mathcal K_v+M^2$ denote the dressed vector operator. The block determinant of the quadratic Hessian gives, after omitting the background-independent matter determinant,}
\begin{align}
{
\Gamma_{K,b}^{(1)}}
&{=
\frac{i}{2}\operatorname{Tr}\ln\left[
(\mathcal A-M^2\Pi_0)\mathcal A^{-1}
\right]}
\nonumber\\
&{=
\frac{i}{2}\operatorname{Tr}\ln\left(
1-\mathcal A^{-1}M^2\Pi_0
\right).}
\label{mixedSchurDeterminant}
\end{align}
{The minus sign in Eq.~\eqref{mixedSchurDeterminant} is the Schur-complement sign associated with the pair of off-diagonal mixing terms; neither individual mixed vertex carries a negative Hessian entry. Using}
\begin{equation}
{
\mathcal A^{-1}
=\frac{\Pi_{1/2}}{\Box+\Omega^2}+\frac{\Pi_0}{\Omega^2},
\qquad
\Omega^2=m^2+M^2,
}
\end{equation}
{one obtains}
\begin{equation}
{
\mathcal A^{-1}M^2\Pi_0
=\frac{M^2}{\Omega^2}\Pi_0
\equiv r\Pi_0.
}
\label{mixedRatioDef}
\end{equation}
{Thus the transverse projector drops out and}
\begin{equation}
{
\Gamma_{K,b}^{(1)}
=\frac{i}{2}\operatorname{Tr}\left[
\Pi_0\ln(1-r)
\right].
}
\label{mixedProjectedDeterminant}
\end{equation}

{The same expression follows from the closed-ring expansion. Since the dressed two-point kernel satisfies $G_D=-\mathcal A^{-1}\delta^8$, the repeated supergraph link is}
\begin{equation}
{
\mathcal Y\equiv M^2\Pi_0G_D=-r\Pi_0.
}
\end{equation}
{The complete ring series is therefore}
\begin{equation}
{
\Gamma_{K,b}^{(1)}
=\frac{i}{2}\sum_{n=1}^{\infty}
\frac{(-1)^{n+1}}{n}
\operatorname{Tr}(\mathcal Y^n).
}
\label{mixedRingSeries}
\end{equation}
{The factor $1/2$ is the real-field Gaussian factor and $1/n$ follows from cyclic invariance. Both chiral orientations are already contained in $\Pi_0$ and must not be counted again. All products in Eq.~\eqref{mixedRingSeries} denote convolutions over intermediate superspace points; every propagator contains its full $\delta^8$ kernel, and all differential operators act before the coincidence limit is taken. Because $\mathcal Y^n=(-1)^nr^n\Pi_0$, the alternating logarithmic coefficient and the negative dressed propagator combine to give}
\begin{align}
{
\Gamma_{K,b}^{(1)}}
&{=
-i\int d^8z_1\,\frac{1}{\Box}
\sum_{n=1}^{\infty}\frac{r^n}{n}
\delta^4(x_1-x_2)\Big|_{x_2=x_1}}
\nonumber\\
&{=
 i\int d^8z_1\,\frac{1}{\Box}\ln(1-r)
\delta^4(x_1-x_2)\Big|_{x_2=x_1},}
\label{mixedRingDA}
\end{align}
{where we used}
\begin{equation}
{
\left.\Pi_0\delta^8(z_1-z_2)\right|_{\theta_1=\theta_2}
=\frac{2}{\Box}\delta^4(x_1-x_2).
}
\end{equation}
{Since $1-r=m^2/\Omega^2$, the exact operator result is}
\begin{equation}
{
\Gamma_{K,b}^{(1)}
=-i\int d^8z_1\,\frac{1}{\Box}
\ln\frac{\Omega^2}{m^2}
\delta^4(x_1-x_2)\Big|_{x_2=x_1}.
}
\label{mixedOperatorResult}
\end{equation}
{With $\Box^{-1}\to-(k^2+i0)^{-1}$, this becomes}
\begin{equation}
{
\Gamma_{K,b}^{(1)}
=i\int d^8z\int\frac{d^4k}{(2\pi)^4}
\frac{1}{k^2+i0}\ln\frac{\Omega^2}{m^2}.
}
\label{mixedMinkowskiResult}
\end{equation}
{After the Wick rotation used in Eq.~\eqref{WickIdentity}, the Euclidean K\"ahlerian density is}
\begin{equation}
{
K_b^{(1)}
=\ln\frac{\Omega^2}{m^2}
\left[
\mu^\epsilon\int\frac{d^dk_E}{(2\pi)^d}\frac{1}{k_E^2}
\right],
\qquad d=4-\epsilon.
}
\label{mixedEuclideanResult}
\end{equation}
{Equation~\eqref{mixedEuclideanResult} is local in the background superfields but proportional to a scaleless loop integral. Only at this stage do we apply dimensional reduction, for which}
\begin{equation}
{
\mu^\epsilon\int\frac{d^dk_E}{(2\pi)^d}\frac{1}{k_E^2}=0,
\qquad
K_{b,\mathrm{DRED}}^{(1)}=0.
}
\label{mixedDREDResult}
\end{equation}
\subsubsection{\revred{Complete minimal result, DRED, and renormalization}}
\label{subsubsec:minimalComplete}

{
The separate DRED result in Eq.~\eqref{mixedDREDResult} must be distinguished
from the algebraic structure of the complete minimal-model determinant. The
Schur complement of the matter block gives the effective vector operator
\begin{align}
\mathcal A_{\rm eff}
&=(\mathcal K_v+M^2)-M^2\Pi_0
\nonumber\\
&=(\Box+\Omega^2)\Pi_{1/2}+m^2\Pi_0.
\label{totalMinimalEffectiveOperator}
\end{align}
Normalizing by the zero-background operator
$\mathcal K_v=(\Box+m^2)\Pi_{1/2}+m^2\Pi_0$, the total one-loop K\"ahlerian
contribution is
\begin{align}
\Gamma_{K,\min}^{(1)}
&=\frac{i}{2}\operatorname{Tr}\ln
\left(\mathcal A_{\rm eff}\mathcal K_v^{-1}\right)
\nonumber\\
&=\frac{i}{2}\operatorname{Tr}
\left[
\Pi_{1/2}\ln\left(1+\frac{M^2}{\Box+m^2}\right)
\right].
\label{totalMinimalDeterminant}
\end{align}
Thus the longitudinal determinant is exactly background independent. In the
sector decomposition, the longitudinal part of the pure-vector contribution
and the mixed determinant satisfy
\begin{equation}
\Gamma_{K,a}^{L}
=+i\int d^8z_1\,\frac{1}{\Box}
\ln\frac{\Omega^2}{m^2}\,
\delta^4(x_1-x_2)\Big|_{x_2=x_1}
=-\Gamma_{K,b}^{(1)}.
\label{totalMinimalLongitudinalCancellation}
\end{equation}
This cancellation precedes regularization and holds whenever the two sectors
are treated with the same regulator and subtraction prescription. In DRED,
the two local terms also vanish separately because each multiplies the same
scaleless integral. These statements are complementary rather than
contradictory.

Using the coincidence identity for $\Pi_{1/2}$, the exact operator result is
\begin{equation}
\Gamma_{K,\min}^{(1)}
=-i\int d^8z_1\,\frac{1}{\Box}
\ln\left(1+\frac{M^2}{\Box+m^2}\right)
\delta^4(x_1-x_2)\Big|_{x_2=x_1}.
\label{totalMinimalOperatorResult}
\end{equation}
Its Minkowski momentum representation is
\begin{equation}
\Gamma_{K,\min}^{(1)}
=i\int d^8z\int\frac{d^4k}{(2\pi)^4}
\frac{1}{k^2+i0}
\ln\frac{k^2-\Omega^2+i0}{k^2-m^2+i0},
\label{totalMinimalMinkowski}
\end{equation}
and the Wick rotation gives
\begin{equation}
K_{\min}^{(1)}
=\mu^\epsilon\int\frac{d^dk_E}{(2\pi)^d}
\frac{1}{k_E^2}
\ln\frac{k_E^2+\Omega^2}{k_E^2+m^2},
\qquad d=4-\epsilon.
\label{totalMinimalEuclidean}
\end{equation}
For a cutoff regulator, the separate longitudinal and mixed pieces contain
opposite quadratic terms,
\begin{equation}
K_a^L=-\frac{\Lambda^2}{16\pi^2}\ln\frac{\Omega^2}{m^2},
\qquad
K_b^{(1)}=+\frac{\Lambda^2}{16\pi^2}\ln\frac{\Omega^2}{m^2},
\end{equation}
which cancel in the total result. Indeed, the combined UV integrand
behaves as
\begin{equation}
\frac{1}{k_E^2}\ln\frac{k_E^2+\Omega^2}{k_E^2+m^2}
=\frac{M^2}{k_E^4}+O(k_E^{-6}),
\end{equation}
so only a logarithmic background-dependent divergence remains.

In DRED, with
$\Delta_{\overline{\mathrm{DR}}}=2/\epsilon-\gamma_E+\ln(4\pi)$,
the normalized unrenormalized result is
\begin{align}
K_{\min}^{(1)}
={}&\frac{1}{16\pi^2}
\left\{
\Omega^2\left[
\Delta_{\overline{\mathrm{DR}}}+2
-\ln\frac{\Omega^2}{\mu^2}
\right]
\right.
\nonumber\\
&\left.\hspace{2.7cm}
-m^2\left[
\Delta_{\overline{\mathrm{DR}}}+2
-\ln\frac{m^2}{\mu^2}
\right]
\right\}.
\label{totalMinimalUnrenormalized}
\end{align}
The complete background-dependent divergence is therefore
\begin{equation}
K_{\min,\mathrm{div}}^{(1)}
=\frac{M^2}{16\pi^2}\Delta_{\overline{\mathrm{DR}}}
=\frac{g^2\Phi\bar\Phi}{16\pi^2}
\Delta_{\overline{\mathrm{DR}}},
\label{totalMinimalDivergence}
\end{equation}
and is canceled by the local K\"ahler counterterm
\begin{equation}
\delta K_{\min}
=-\frac{g^2\Phi\bar\Phi}{16\pi^2}
\Delta_{\overline{\mathrm{DR}}}.
\label{totalMinimalCounterterm}
\end{equation}
Minimal $\overline{\mathrm{DR}}$ subtraction yields
\begin{align}
K_{\min,\overline{\mathrm{DR}}}^{(1)}
={}&\frac{1}{16\pi^2}
\left\{
\Omega^2\left[2-\ln\frac{\Omega^2}{\mu^2}\right]
-m^2\left[2-\ln\frac{m^2}{\mu^2}\right]
\right\}.
\label{totalMinimalRenormalized}
\end{align}
A further finite local wave-function subtraction may remove
$2M^2/(16\pi^2)$. In the corresponding logarithm-only convention,
\begin{equation}
K_{\min,\log}^{(1)}
=-\frac{1}{16\pi^2}
\left[
\Omega^2\ln\frac{\Omega^2}{\mu^2}
-m^2\ln\frac{m^2}{\mu^2}
\right].
\label{totalMinimalLogOnly}
\end{equation}
Finite local polynomial subtractions do not reverse the sign of the
nonpolynomial logarithmic dependence.

The calculation establishes renormalization of the one-loop K\"ahlerian
sector by a local matter wave-function counterterm.
{It is not itself a proof of perturbative renormalizability of
the complete explicit Proca theory; the status of that question at $\alpha=0$
is discussed in Sec.~\ref{subsec:renormScheme}.} Classically, the action used here is the unitary-gauge form of the minimal Stueckelberg realization of Sec.~\ref{subsec:stueckelbergRealization}, so no matching is involved at this level; completions in which additional heavy fields are integrated out can generate further local operators whose Wilson coefficients require matching. For the higher-derivative operator of Sec.~\ref{subsec:generalFormFactor}, the exact full-range polynomial model must likewise be distinguished from a truncated derivative expansion valid for $|\alpha k^2|\ll1$; see Sec.~\ref{subsec:renormScheme}. Moreover, the logarithms above are local functions
of the background superfields within the K\"ahlerian approximation; the
full derivative-dependent and momentum-nonlocal effective action has not been
calculated.
}

\subsection{\texorpdfstring{\revred{General transverse form factor $f(\Box)$}}{General transverse form factor f(Box)}}
\label{subsec:generalFormFactor}

{All background-field, combinatorial, full-kernel, projector,
Fourier, Wick-rotation, and DRED conventions were established once in
Subsections~\ref{subsec:kahlerScope}--\ref{subsec:commonDeterminant}. We
therefore present only the operator replacements and analytic features that
are specific to the general transverse form factor.}

\subsubsection{\revred{General vector operators and propagators}}
\label{subsubsec:generalOperators}

{
We generalize the transverse kinetic operator while retaining the explicit
Proca mass in both projector sectors. Let
\begin{equation}
\mathcal K_f
=f(\Box)\Pi_{1/2}+m^2
=\bigl[f(\Box)+m^2\bigr]\Pi_{1/2}+m^2\Pi_0 .
\label{generalFOperator}
\end{equation}
The quadratic vector action must contain the real vector superfield on both
sides of the operator:
\begin{equation}
S_{\mathrm{vec},f}
=\frac12\int d^8z\,
v\,\mathcal K_f\,v
=\frac12\int d^8z\,
v\left\{\bigl[f(\Box)+m^2\bigr]\Pi_{1/2}
+m^2\Pi_0\right\}v .
\label{generalFVectorAction}
\end{equation}
We assume that \(f(z)=z\,h(z)\), where \(h\) is real analytic near the
origin and \(h(0)=1\). The additional analytic and invertibility assumptions
needed for the Wick rotation are stated below.

With the same background split used in the minimal model, the complete
quadratic action is
\begin{align}
S_f^{(2)}
={}&
\frac12\int d^8z\,v\bigl(\mathcal K_f+M^2\bigr)v
+\int d^8z\,\bar\varphi\varphi
\nonumber\\
&+
g\int d^8z\,
\bigl(\bar\Phi\,v\,\varphi+\bar\varphi\,v\,\Phi\bigr),
\qquad
M^2=g^2\Phi\bar\Phi .
\label{generalFQuadraticAction}
\end{align}
Thus the background-dependent vector insertion is again \(V_{vv}=M^2\),
with no residual factor \(1/2\).

Projector orthogonality gives the free propagator
\begin{equation}
\langle v(z_1)v(z_2)\rangle_{0,f}
=
-\left[
\frac{\Pi_{1/2}}{f(\Box)+m^2}
+\frac{\Pi_0}{m^2}
\right]\delta^8(z_1-z_2).
\label{generalFFreePropagator}
\end{equation}
Equivalently,
\begin{equation}
\mathcal K_f^{-1}
=
\frac{1}{f(\Box)+m^2}
\left[
1+\frac{f(\Box)}{m^2}\Pi_0
\right].
\label{generalFFreeInverseAlternative}
\end{equation}
Since \(f(\Box)/\Box=h(\Box)\) is regular, the inverse powers of \(\Box\)
inside the projectors do not by themselves generate an additional massless
pole. The transverse poles are determined by
\(f(-k^2)+m^2=0\), while the longitudinal sector is algebraic.

{The relation with the gauge-invariant higher-derivative SQED model of Ref.~\cite{HDgauge} can be stated explicitly. In the limit $m^2=0$, take $f(\Box)=\Box+\alpha\Box^2=\alpha\Box(\Box+\Lambda_{\mathrm{HD}}^2)$ and denote the vector superfield and coupling used in Ref.~\cite{HDgauge} by $v_{\rm HD}$ and $g_{\rm HD}$, respectively. The required identifications are}
\[
{\Lambda_{\mathrm{HD}}^2=\frac{1}{\alpha},\qquad v_{\rm HD}=\sqrt{\alpha}\,v,\qquad g_{\rm HD}=\frac{g}{\sqrt{\alpha}}.}
\]
{With these definitions, the transverse quadratic term becomes}
\[
{\frac{1}{2}\int d^8z\,v\Pi_{1/2}f(\Box)v
=\frac{1}{2}\int d^8z\,v_{\rm HD}\Pi_{1/2}\Box(\Box+\Lambda_{\mathrm{HD}}^2)v_{\rm HD},}
\]
{and the exponential interaction preserves its form, $g v=g_{\rm HD}v_{\rm HD}$. Hence $[v]=0$ and $[g]=0$ in the present normalization, whereas $[v_{\rm HD}]=-1$ and $[g_{\rm HD}]=1$ in the normalization of Ref.~\cite{HDgauge}; the apparent mismatch between the mass dimensions of the two propagators is therefore a consequence of the field normalization. At $m^2=0$, however, Abelian supergauge invariance is restored and the longitudinal quadratic operator is no longer invertible. A gauge-fixing term must consequently be introduced before the full propagators can be compared. The mapping above applies directly to the transverse sector, while the longitudinal sector of Ref.~\cite{HDgauge} is gauge-fixing dependent.}

Defining \(\Omega^2=m^2+M^2\), the dressed vector operator and propagator are
\begin{align}
\mathcal A_f
&\equiv\mathcal K_f+M^2
=\bigl[f(\Box)+\Omega^2\bigr]\Pi_{1/2}
+\Omega^2\Pi_0 ,
\label{generalFDressedOperator}
\\
\langle v(z_1)v(z_2)\rangle_{D,f}
&=
-\left[
\frac{\Pi_{1/2}}{f(\Box)+\Omega^2}
+\frac{\Pi_0}{\Omega^2}
\right]\delta^8(z_1-z_2).
\label{generalFDressedPropagator}
\end{align}
In particular,
\begin{equation}
\Pi_0\langle v(z_1)v(z_2)\rangle_{D,f}
=
-\frac{\Pi_0}{\Omega^2}\delta^8(z_1-z_2),
\label{generalFDressedLongitudinal}
\end{equation}
which is independent of \(f\).
}

\subsubsection{\revred{Pure-vector and mixed determinants}}
\label{subsubsec:generalDeterminants}

{
The pure-vector contribution is still normalized by the determinant at
$M^2=0$,
\begin{equation}
\Gamma_{K,a}^{(1)}[f]
=\frac{i}{2}\operatorname{Tr}\ln\left[
(\mathcal K_f+M^2)\mathcal K_f^{-1}
\right]
=\frac{i}{2}\operatorname{Tr}\ln\left(
1+M^2\mathcal K_f^{-1}
\right).
\label{generalFPureVectorDeterminant}
\end{equation}
The ring factors, full superspace kernels, and projector coincidence
identities have already been established in
Subsections~\ref{subsec:backgroundHessian} and~\ref{subsec:commonDeterminant}.
Applying those common steps with the replacement
$\Box\mapsto f(\Box)$ only in the transverse sector gives directly
\begin{align}
\Gamma_{K,a}^{(1)}[f]
={}&-i\int d^8z_1\,\frac{1}{\Box}
\left[
\ln\left(1+\frac{M^2}{f(\Box)+m^2}\right)
-\ln\left(1+\frac{M^2}{m^2}\right)
\right]
\nonumber\\[-0.2em]
&\hspace{7.0cm}\times
\delta^4(x_1-x_2)\Big|_{x_2=x_1}.
\label{generalFPureVectorResummed}
\end{align}

The matter bridge is unchanged, $\Sigma_\chi=M^2\Pi_0\delta^8$, and the
longitudinal projection in Eq.~\eqref{generalFDressedLongitudinal} is
independent of $f$. Consequently, the mixed determinant is the same local
operator found in the minimal model,
\begin{equation}
\Gamma_{K,b}^{(1)}[f]
=-i\int d^8z_1\,\frac{1}{\Box}
\ln\frac{\Omega^2}{m^2}
\delta^4(x_1-x_2)\Big|_{x_2=x_1},
\label{generalFMixedOperatorResult}
\end{equation}
and, separately,
\begin{equation}
K_b^{(1)}[f]
=\ln\frac{\Omega^2}{m^2}
\left[\mu^\epsilon\int\frac{d^dk_E}{(2\pi)^d}\frac{1}{k_E^2}\right]
=0
\qquad\text{in DRED}.
\label{generalFMixedEuclidean}
\end{equation}
This isolated vanishing is regulator dependent, whereas its cancellation
against the longitudinal pure-vector term is algebraic. Indeed, the Schur
complement gives
\begin{equation}
\mathcal A_{f,\mathrm{eff}}
=\bigl[f(\Box)+\Omega^2\bigr]\Pi_{1/2}+m^2\Pi_0,
\label{generalFEffectiveOperator}
\end{equation}
and hence the complete determinant is purely transverse,
\begin{equation}
\Gamma_{K,\mathrm{tot}}^{(1)}[f]
=\frac{i}{2}\operatorname{Tr}
\left[
\Pi_{1/2}\ln\left(1+\frac{M^2}{f(\Box)+m^2}\right)
\right].
\label{generalFTotalDeterminant}
\end{equation}
}

{Setting $f(\Box)=\Box$ is a direct algebraic choice and should be distinguished from the singular higher-derivative limit discussed below:}
\begin{equation}
{
f(\Box)=\Box
\quad\Longrightarrow\quad
\Gamma_{K,\mathrm{tot}}^{(1)}[f]
=\Gamma_{K,\min}^{(1)}.}
\label{generalFMinimalSpecialization}
\end{equation}

{
With $\Box\to-k^2$ and a common causal prescription for every zero of
$f(-k^2)+s$, the pure-vector contribution is
\begin{equation}
\Gamma_{K,a}^{(1)}[f]
=i\int d^8z\int\frac{d^4k}{(2\pi)^4}\frac{1}{k^2+i0}
\left[
\ln\frac{f(-k^2)+\Omega^2}{f(-k^2)+m^2}
-\ln\frac{\Omega^2}{m^2}
\right]_{\!\mathrm{causal}}.
\label{generalFPureVectorMinkowski}
\end{equation}
The Wick rotation and DRED prescription fixed in the minimal calculation give
\begin{equation}
K_{a,\mathrm{DRED}}^{(1)}[f]
=\mu^\epsilon\int\frac{d^dk_E}{(2\pi)^d}\frac{1}{k_E^2}
\ln\frac{f(k_E^2)+\Omega^2}{f(k_E^2)+m^2},
\label{generalFPureVectorDRED}
\end{equation}
because the background-independent second logarithm multiplies a scaleless
integral. After the algebraic longitudinal cancellation, the same expression
is the complete Euclidean density,
\begin{equation}
K_{\mathrm{tot}}^{(1)}[f]
=\mu^\epsilon\int\frac{d^dk_E}{(2\pi)^d}\frac{1}{k_E^2}
\ln\frac{f(k_E^2)+\Omega^2}{f(k_E^2)+m^2}.
\label{generalFTotalEuclidean}
\end{equation}
Here $f_E(k_E^2)\equiv f(k_E^2)$. This continuation assumes that $f$ is
analytic in the region swept by the Wick rotation, that no pole or cut crosses
the contour, that $f(\Box)+m^2$ and $f(\Box)+\Omega^2$ are invertible with the
prescribed boundary conditions, and that the logarithm is assigned a continuous
branch. A manifestly real Euclidean expression follows, in particular, when
$f(x)+m^2>0$ and $f(x)+\Omega^2>0$ for $x\geq0$.
}

\subsubsection{\texorpdfstring{\revred{Case $f(\Box)=\Box+\alpha\Box^2$}}{Case f(Box)=Box+alpha Box squared}}
\label{subsubsec:polynomialSpecialization}

{
For $f(\Box)=\Box+\alpha\Box^2$, define
\begin{equation}
\lambda_\pm(s)
=\frac{1\pm\sqrt{1-4\alpha s}}{2\alpha},
\qquad
\alpha k_E^4+k_E^2+s
=\alpha\bigl(k_E^2+\lambda_+(s)\bigr)
    \bigl(k_E^2+\lambda_-(s)\bigr),
\label{generalFPolynomialRoots}
\end{equation}
where the square-root branch is chosen continuously from
$\sqrt{1-4\alpha s}\to1$ as $\alpha\to0$. The polynomial integral is
\begin{equation}
K_{a,\mathrm{DRED}}^{(1)}(\alpha)
=\mu^\epsilon\int\frac{d^dk_E}{(2\pi)^d}\frac{1}{k_E^2}
\ln\frac{\alpha k_E^4+k_E^2+\Omega^2}
{\alpha k_E^4+k_E^2+m^2}.
\label{generalFPolynomialIntegral}
\end{equation}
Using
\begin{equation}
\mathcal J(A,B)
=\frac{1}{16\pi^2}
\left\{
A\left[\Delta_{\overline{\mathrm{DR}}}+2-\ln\frac{A}{\mu^2}\right]
-B\left[\Delta_{\overline{\mathrm{DR}}}+2-\ln\frac{B}{\mu^2}\right]
\right\},
\label{generalFMasterLogIntegral}
\end{equation}
one obtains
\begin{align}
K_{a,\mathrm{DRED}}^{(1)}(\alpha)
=\frac{1}{16\pi^2}\sum_{\sigma=\pm}
\bigg\{&
\lambda_\sigma(\Omega^2)
\left[\Delta_{\overline{\mathrm{DR}}}+2
-\ln\frac{\lambda_\sigma(\Omega^2)}{\mu^2}\right]
\nonumber\\
&-\lambda_\sigma(m^2)
\left[\Delta_{\overline{\mathrm{DR}}}+2
-\ln\frac{\lambda_\sigma(m^2)}{\mu^2}\right]
\bigg\}.
\label{generalFPolynomialDREDResult}
\end{align}
Since
\begin{equation}
\lambda_+(s)+\lambda_-(s)=\frac1\alpha,
\qquad
\lambda_+(s)\lambda_-(s)=\frac{s}{\alpha},
\label{generalFRootIdentities}
\end{equation}
the coefficients of $\Delta_{\overline{\mathrm{DR}}}$, the finite constant,
and $\ln\mu^2$ cancel between $s=\Omega^2$ and $s=m^2$. The normalized
field-dependent result is therefore finite for fixed nonzero $\alpha$:
\begin{align}
K_a^{(1)}(\alpha)
=-\frac{1}{16\pi^2}\sum_{\sigma=\pm}
\bigg[&
\lambda_\sigma(\Omega^2)
\ln\frac{\lambda_\sigma(\Omega^2)}{\mu^2}
\nonumber\\
&-\lambda_\sigma(m^2)
\ln\frac{\lambda_\sigma(m^2)}{\mu^2}
\bigg].
\label{generalFPolynomialFiniteResult}
\end{align}
The scale $\mu$ cancels in the complete expression, and
$K_a^{(1)}(M^2=0)=0$ is manifest. As an independent check,
\begin{equation}
\frac{\partial K_a^{(1)}}{\partial s}
=
\frac{1}{16\pi^2\sqrt{1-4\alpha s}}
\ln\frac{\lambda_+(s)}{\lambda_-(s)},
\label{generalFPolynomialDerivativeCheck}
\end{equation}
which is positive in the real domain $\alpha>0$, $0<4\alpha s<1$ and fixes
both the sign of the Euclidean integral and the factor $1/(16\pi^2)$.
}

\subsubsection{\texorpdfstring{Roots, poles, and analytic domain}{Roots, poles, and analytic domain}}
\label{subsubsec:rootsPoles}

{
To characterize the analytic domain of \eqref{generalFPolynomialFiniteResult}, let
\begin{equation}
\Delta_s\equiv1-4\alpha s,
\qquad s\in\{m^2,\Omega^2\}.
\label{hdDiscriminant}
\end{equation}
For $\alpha>0$ and $0<4\alpha s<1$, both parameters
$\lambda_\pm(s)$ in \eqref{generalFPolynomialRoots} are real and positive,
with $\lambda_-(s)<\lambda_+(s)$. On a real bosonic background,
the vacuum and dressed operators therefore possess two real positive
transverse poles whenever
\begin{equation}
4\alpha\Omega^2\leq1,
\qquad
0\leq M^2\leq\frac{1}{4\alpha}-m^2.
\label{hdRealPoleBackgroundDomain}
\end{equation}
The bound on $M^2$ assumes $M^2\geq0$ and also requires
$4\alpha m^2\leq1$.

The Minkowski transverse denominator is
\begin{equation}
f(-k^2)+s
=\alpha(k^2-\lambda_+(s))(k^2-\lambda_-(s)).
\label{hdMinkowskiFactorization}
\end{equation}
Accordingly, the scalar coefficient of the transverse propagator can be
written, for nondegenerate roots, as
\begin{equation}
-\frac{1}{f(-k^2)+s}
=\frac{1}{\sqrt{\Delta_s}}
\left[
\frac{1}{k^2-\lambda_-(s)}
-\frac{1}{k^2-\lambda_+(s)}
\right].
\label{hdPoleDecomposition}
\end{equation}
The corresponding residues are
\begin{equation}
Z_-(s)=\frac{1}{\sqrt{\Delta_s}},
\qquad
Z_+(s)=-\frac{1}{\sqrt{\Delta_s}}.
\label{hdPoleResidues}
\end{equation}
For small positive $\alpha$,
\begin{equation}
\lambda_-(s)=s+\alpha s^2+O(\alpha^2),
\qquad
\lambda_+(s)=\frac1\alpha-s+O(\alpha),
\label{hdPoleSmallAlpha}
\end{equation}
so the light pole is continuously connected to the Proca pole, whereas
the additional higher-derivative pole has the opposite residue. Thus, under standard quantization, a
fundamental two-pole interpretation contains a ghostlike transverse
multiplet; supersymmetry does not remove this relative residue sign.
These statements refer to the exact polynomial model; in a truncated derivative expansion valid for $|\alpha k^2|\ll1$, an apparent additional pole at the scale $|k^2|\sim1/|\alpha|$ lies outside the controlled domain and is not a prediction (Sec.~\ref{subsec:renormScheme}).

At $\Delta_s=0$, the poles coalesce at
$\lambda_0=1/(2\alpha)$ and
\begin{equation}
-\frac{1}{f(-k^2)+s}
=-\frac{1}{\alpha(k^2-\lambda_0)^2}.
\label{hdDegeneratePole}
\end{equation}
Although the propagator then has a double pole, the one-loop K\"ahlerian
combination remains regular. In particular,
\begin{equation}
\lim_{\Delta_s\to0}
\frac{1}{16\pi^2\sqrt{\Delta_s}}
\ln\frac{1+\sqrt{\Delta_s}}{1-\sqrt{\Delta_s}}
=\frac{1}{8\pi^2}.
\label{hdDegeneratePotentialLimit}
\end{equation}

For $\alpha>0$ and $4\alpha s>1$, the pole parameters become the
complex-conjugate pair
\begin{equation}
\lambda_\pm(s)=\frac{1\pm i\rho_s}{2\alpha},
\qquad
\rho_s=\sqrt{4\alpha s-1}.
\label{hdComplexPoles}
\end{equation}
This does not make the Euclidean integral singular: the original polynomial
$\alpha k_E^4+k_E^2+s$ is strictly positive for $k_E^2\geq0$.
Choosing conjugate branches for the logarithms gives the explicitly real
continuation
\begin{equation}
\sum_{\sigma=\pm}\lambda_\sigma(s)
\ln\frac{\lambda_\sigma(s)}{\mu^2}
=\frac{1}{\alpha}
\left[
\ln\frac{\sqrt{s/\alpha}}{\mu^2}
-\rho_s\arctan\rho_s
\right].
\label{hdComplexRootContinuation}
\end{equation}
Hence \eqref{generalFPolynomialFiniteResult} remains real and continuous
for positive $\alpha$, even beyond the two-real-pole domain. The poles no
longer describe ordinary asymptotic particles there.

For $\alpha<0$, one of the $\lambda_\pm$ is negative. The Euclidean
polynomial then has a zero at positive $k_E^2$, obstructing the standard
Wick-rotated integral, while the additional Minkowski pole has negative
mass squared and negative residue. We therefore restrict the unqualified
Euclidean formulation to $\alpha>0$. A negative value of $\alpha$ could
only be used with an explicit cutoff below that zero and a separate contour
prescription.}

\subsubsection{\texorpdfstring{\revred{Divergence and the $\alpha\to0$ limit}}{Divergence and the alpha to 0 limit}}
\label{subsubsec:alphaLimit}

{
The UV structure and the minimal-theory limit can now be made
explicit. Before normalizing the determinant at vanishing matter
background, define the single-polynomial contribution
\begin{equation}
\mathcal U_\alpha(s)
\equiv
\mu^\epsilon\int\frac{d^d k_E}{(2\pi)^d}\frac{1}{k_E^2}
\ln\bigl(\alpha k_E^4+k_E^2+s\bigr),
\qquad s\in\{m^2,\Omega^2\}.
\label{hdSinglePolynomialContribution}
\end{equation}
An arbitrary constant inside the logarithm is immaterial because it
multiplies the scaleless integral $\int_k k_E^{-2}$. Using the roots in
Eq.~\eqref{generalFPolynomialRoots}, one obtains
\begin{equation}
\mathcal U_\alpha(s)
=\frac{1}{16\pi^2}
\sum_{\sigma=\pm}\lambda_\sigma(s)
\left[
\Delta_{\overline{\mathrm{DR}}}+2
-\ln\frac{\lambda_\sigma(s)}{\mu^2}
\right].
\label{hdSinglePolynomialDRED}
\end{equation}
Since $\lambda_+(s)+\lambda_-(s)=1/\alpha$, its complete pole is
\begin{equation}
\mathcal U_{\alpha,\mathrm{div}}(s)
=\frac{1}{16\pi^2\alpha}
\Delta_{\overline{\mathrm{DR}}}.
\label{hdRawFieldIndependentDivergence}
\end{equation}
This divergence is independent of $s$ and hence of the matter background.
It is a background-independent vacuum-functional normalization term. In
rigid flat superspace it is also a constant shift of the K\"ahler
potential, whose full-superspace integral vanishes. If the unnormalized
vacuum functional is retained, the pole may formally be removed by the
field-independent subtraction
$\delta K_{\rm vac}^{\rm HD}
=-\Delta_{\overline{\mathrm{DR}}}/(16\pi^2\alpha)$.

The K\"ahlerian correction used here is normalized at $M^2=0$. In the polynomial case we denote the complete normalized density by
$K_{\rm HD}^{(1)}(\alpha)\equiv K_a^{(1)}(\alpha)$; after the longitudinal cancellation it is also the form of $K_{\mathrm{tot}}^{(1)}[f]$ for $f(\Box)=\Box+\alpha\Box^2$. Thus,
\begin{equation}
K_{\rm HD}^{(1)}(\alpha)
=\mathcal U_\alpha(\Omega^2)-\mathcal U_\alpha(m^2),
\qquad
K_{\rm HD,div}^{(1)}(\alpha)=0
\quad(\alpha\ne0).
\label{hdNormalizedDivergenceCancellation}
\end{equation}
Thus no background-dependent one-loop K\"ahler counterterm is required
for fixed nonzero $\alpha$ in the exact full-range polynomial model. In particular, all terms of order
$\alpha^{-1}$, including the pole and the leading heavy-root contribution,
cancel from the normalized result.

The small-$\alpha$ behavior follows from
\begin{equation}
\lambda_-(s)=s+\alpha s^2+2\alpha^2s^3+O(\alpha^3),
\qquad
\lambda_+(s)=\frac1\alpha-s-\alpha s^2-2\alpha^2s^3+O(\alpha^3).
\label{hdRootsSmallAlphaExtended}
\end{equation}
Substitution into Eq.~\eqref{generalFPolynomialFiniteResult} gives
\begin{align}
K_{\rm HD}^{(1)}(\alpha)
=\frac{1}{16\pi^2}\bigg\{&
M^2\left[\ln\frac{1}{\alpha m^2}+1\right]
-\Omega^2\ln\frac{\Omega^2}{m^2}
\nonumber\\
&+\alpha\left[
\Omega^4\left(\ln\frac{1}{\alpha\Omega^2}-\frac12\right)
-m^4\left(\ln\frac{1}{\alpha m^2}-\frac12\right)
\right]
+O(\alpha^2\ln\alpha)
\bigg\}.
\label{hdSmallAlphaExpansion}
\end{align}
There is no power-law $\alpha^{-1}$ term in the normalized expression. Its first singular term is the local threshold logarithm
$M^2\ln(1/\alpha)/(16\pi^2)$.

The origin of this nonuniform limit is transparent in the radial form
\begin{equation}
K_{\rm HD}^{(1)}(\alpha)
=\frac{1}{16\pi^2}\int_0^\infty d x\,
\ln\frac{\alpha x^2+x+\Omega^2}
    {\alpha x^2+x+m^2},
\qquad x=k_E^2.
\label{hdRadialIntegral}
\end{equation}
At fixed $x$ the integrand tends to the minimal Proca integrand. However,
in the expanding region
$m^2,\Omega^2\ll x\ll1/\alpha$ it behaves as $M^2/x$, whereas for
$x\gg1/\alpha$ it behaves as $M^2/(\alpha x^2)$. The moving scale
$x\simeq1/\alpha$ therefore cuts off the minimal logarithmic UV
region and generates the term $M^2\ln(1/\alpha)/(16\pi^2)$. The limit
$\alpha\to0$ cannot be interchanged with the loop integration.

More precisely, comparison with the minimal unrenormalized result shows
that the higher-derivative threshold implements the correspondence
\begin{equation}
\Delta_{\overline{\mathrm{DR}}}
\longleftrightarrow
\ln\frac{1}{\alpha\mu^2}-1.
\label{hdMinimalDivergenceCorrespondence}
\end{equation}
The $\overline{\mathrm{DR}}$ result of the minimal model is recovered after the local
matching subtraction
\begin{equation}
\delta K_{\rm match}(\alpha,\mu)
=-\frac{M^2}{16\pi^2}
\left[\ln\frac{1}{\alpha\mu^2}-1\right],
\qquad
\lim_{\alpha\to0^+}
\left[K_{\rm HD}^{(1)}+\delta K_{\rm match}\right]
=K_{\min,\overline{\mathrm{DR}}}^{(1)}.
\label{hdMinimalMatchingLimit}
\end{equation}
The matching term is local and proportional to
$M^2=g^2\Phi\bar\Phi$; it represents threshold matching of the matter
K\"ahler normalization, not an additional UV pole of the
fixed-$\alpha$ theory; its status in a truncated derivative expansion is discussed in Sec.~\ref{subsec:renormScheme}. Hence the field-independent divergence at
nonzero $\alpha$ and the field-dependent divergence of the minimal model
are compatible: UV integration and the singular limit
$\alpha\to0$ are noncommuting operations.
}

\section{\revgreen{Discussion and physical scope}}
\label{sec:discussionScope}

\subsection{\revgreen{Renormalization and scheme dependence}}
\label{subsec:renormScheme}

{The minimal result in Eqs.~\eqref{totalMinimalDivergence}--\eqref{totalMinimalRenormalized} establishes a limited but definite renormalization statement. The complete one-loop K\"ahlerian divergence is local and proportional to $g^2\Phi\bar\Phi$; it is absorbed by the matter K\"ahler normalization. Minimal $\overline{\mathrm{DR}}$ subtraction removes $\Delta_{\overline{\mathrm{DR}}}$, whereas additional finite local subtractions correspond to a choice of normalization condition. They may change polynomial terms in the background but not the sign of the nonpolynomial logarithmic dependence. {The renormalizability of the complete theory is a separate question, taken up next.} The effective-theory status of the model is formulated in Sec.~\ref{subsec:scopeStueckelberg} in terms of its minimal Stueckelberg realization and of possible non-minimal completions.}

{At $\alpha=0$ a closely related model---the minimal Abelian $\mathcal N=1$ Stueckelberg theory with conserved, vector-like and hence anomaly-free matter---was studied by Delbourgo~\cite{Delbourgo:1975sst}, who gave a superspace supergraph power-counting argument for perturbative renormalizability and identified the corresponding counterterm structure.} {There the vector superfield is renormalized multiplicatively, so that the mass and coupling renormalizations are purely multiplicative, and the counterterm shown to be absent is the independent chiral-matter mass counterterm.} {Under the assumptions of an anomaly-free minimal Stueckelberg formulation of the type discussed in Sec.~\ref{subsec:stueckelbergRealization}, physical on-shell amplitudes may be renormalized through the gauge-fixed parent formulation.} {Related non-supersymmetric connections between the Proca and Stueckelberg generating functionals through finite field-dependent BRST transformations are discussed in Refs.~\cite{Upadhyay:2012ffb,Pal:2026ffb}.} {This is not a literal all-order theorem for the single-chiral-field model used in the present calculation, and no corresponding conclusion is established for the higher-derivative model with $\alpha\neq0$.}

{A separate question concerns off-shell quantities, where two non-supersymmetric analyses serve here only as analogies: in massive QED, Sonoda found additional momentum-space contact divergences and a noncommutativity between the unitary-gauge limit and minimal subtraction~\cite{Sonoda:2001mqed}, and in the Abelian Higgs model he analyzed related singularities associated with composite interpolating fields~\cite{Sonoda:2001higgs}, the physical $S$-matrix being unaffected in both cases.} {The K\"ahlerian divergence obtained above is not identified with those coincident-point divergences.} {It is absorbed independently by the local matter K\"ahler wave-function counterterm, the finite local remainder is scheme dependent, and an on-shell normalization condition would require derivative-dependent two-point information, pole residues and, in general, a physical amplitude, none of which is computed here.} {In the present calculation, supersymmetry organizes the divergence as a local full-superspace K\"ahler counterterm, namely chiral-superfield wave-function renormalization, while DRED with the $\overline{\mathrm{DR}}$ prescription preserves this supersymmetric organization at one loop; supersymmetry does not by itself remove the general off-shell subtleties of unitary-gauge variables.}

{For fixed nonzero $\alpha$, the exact polynomial model, integrated over the full Euclidean range, gives a finite normalized higher-derivative correction; the pole in Eq.~\eqref{hdRawFieldIndependentDivergence} is independent of the matter background and cancels under the normalization in Eq.~\eqref{hdNormalizedDivergenceCancellation}. The minimal model is not obtained by a naive integrated limit: the heavy root generates a local threshold logarithm, and Eq.~\eqref{hdMinimalMatchingLimit} shows how a local subtraction recovers the $\overline{\mathrm{DR}}$ result of the minimal model. If $\alpha\Box^2$ is instead regarded as the leading term of a derivative expansion valid for $|\alpha k^2|\ll1$, finiteness at fixed $\alpha$ is a property of the exact full-range polynomial model rather than of the truncated theory, whose local operator basis may require additional counterterms at higher orders in $\alpha$. In that reading the additional pole and the region $k_E^2\sim1/|\alpha|$ lie outside the controlled domain and depend on the UV completion, whereas the complete local term in Eq.~\eqref{hdMinimalMatchingLimit} is completion-dependent UV/threshold matching information; only its $\mu$ dependence is fixed by the low-energy renormalization group.}

\subsection{\revgreen{Mixed determinant and regulator dependence}}

{The mixed sector illustrates the distinction between an algebraic identity and a regulator property. Considered separately, the mixed determinant is a local scaleless tadpole and therefore vanishes in DRED, as in Eq.~\eqref{mixedDREDResult}. In the complete determinant it cancels the pure-vector longitudinal term before regularization, Eq.~\eqref{totalMinimalLongitudinalCancellation}. A common cutoff gives equal and opposite local quadratic terms. The cancellation is a property of the specific Hessian and coupling considered here and should not be interpreted as a universal signature of supersymmetric massive-vector dynamics.}

\subsection{\revgreen{Relation to previous work and scope of novelty}}
\label{subsec:noveltyScope}

{The diagrammatic organization into pure-vector rings and mixed rings, their $1/(2n)$ symmetry factors, and their logarithmic resummation belong to the established superspace effective-action framework. Likewise, higher-derivative Abelian gauge kinetic operators and one-loop K\"ahlerian supergraphs were studied previously in Ref.~\cite{HDgauge}. The novelty claimed here is therefore not a new resummation technique.}

{The first model-specific result is the complete one-loop K\"ahlerian determinant of the stand-alone supersymmetric Proca interaction used in this paper. In the Proca variables the longitudinal projector sector is not generated by a gauge-fixing functional; it is the unitary-gauge remnant of the Stueckelberg mass term (Sec.~\ref{subsec:stueckelbergRealization}). The calculation shows directly, both from the Schur complement and from the closed supergraph rings, that the matter bridge removes the background dependence of this nonpropagating longitudinal block and leaves a purely transverse determinant. This algebraic statement is stronger than, and conceptually distinct from, setting the isolated mixed tadpole to zero in DRED.}

{The second model-specific result is the extension to the operator $\mathcal K_f=f(\Box)\Pi_{1/2}+m^2$ with a nonzero Proca mass retained in the longitudinal sector. This simultaneous presence of a general transverse form factor and an algebraic Proca block is not obtained by a literal substitution in the gauge-invariant theory of Ref.~\cite{HDgauge}. The two formulations meet only in the transverse, $m^2\to0$ limit after the field normalization described in Sec.~\ref{subsec:generalFormFactor}; their longitudinal sectors differ, because in Ref.~\cite{HDgauge} the longitudinal operator is fixed by a gauge-fixing functional, whereas here it is the unitary-gauge Stueckelberg mass remnant and the Proca variables carry no residual gauge redundancy.}

{The third result is the corrected and fully analyzed quadratic example $f(\Box)=\Box+\alpha\Box^2$. Besides the exact factor $1/(16\pi^2)$ and the normalized sign, the analysis includes the background-dependent discriminant, analytic continuation through complex-conjugate roots, pole residues and the double-pole limit, the field-independent $\alpha^{-1}$ divergence of the unnormalized determinant, and the nonuniform $\alpha\to0$ matching to the $\overline{\mathrm{DR}}$ result of the minimal model. These results characterize the exact polynomial model; their reading within a truncated derivative expansion is given in Sec.~\ref{subsec:renormScheme}.}

{Taken together, the minimal determinant, the general-$f$ operator identity, and the spectral and matching analysis of the quadratic form factor provide explicit one-loop results for a Proca action that is classically the unitary-gauge form of the minimal supersymmetric Stueckelberg realization, a setting in which the complete real vector superfield and its algebraic longitudinal block must be retained. The earlier three-dimensional massive-super-QED calculation of Ref.~\cite{mass3} is part of the same research program, but it concerns different dimensionality and dynamics and does not supply the four-dimensional K\"ahlerian results obtained here.}

{The results established in this paper are the propagator and dynamical constraint of the explicit Proca superfield, the one-loop K\"ahlerian determinants, the renormalization of the minimal K\"ahlerian correction in the $\overline{\mathrm{DR}}$ scheme with DRED as the regulator, and the pole and matching structure of the quadratic higher-derivative example.}

\subsection{\revgreen{Model scope, constraints, and spectral interpretation}}
\label{subsec:scopeStueckelberg}

{The calculated Proca action is the exact classical unitary-gauge form of the minimal Abelian $\mathcal N=1$ Stueckelberg realization of Sec.~\ref{subsec:stueckelbergRealization}; at the quantum level, a consistent gauge-invariant parent requires anomaly-free matter. The effective-theory content is therefore organized by the completion, not by the gauge-broken Proca symmetry alone. In the minimal realization, the classical action is Eq.~\eqref{stueckelbergToProca} and no matching is involved; completions containing additional heavy fields can generate, after these fields are integrated out, further local vector-sector and matter operators whose Wilson coefficients require matching to the specified completion. The K\"ahlerian calculation has a correspondingly defined scope: it determines the leading derivative-free full-superspace quantum correction of the minimal realization and supplies a benchmark for comparison with such completions. {The renormalizability statement inferred from the Stueckelberg formulation requires an anomaly-free completion of the matter sector; this requirement does not alter the one-loop background-dependent K\"ahler result in the single-background sector studied here.}}

{The real-linear constraint and its vector Lorenz condition follow dynamically from the Proca field equation. At one loop, matter corrections to the vector two-point function are transverse, so this constraint persists for nonzero renormalized mass; an all-order statement would require an anomaly-free formulation and a separate analysis of the corresponding quantum Ward identity. The minimal free spin-one Hamiltonian is positive for $m^2>0$, but supersymmetry alone does not establish interacting spectral positivity. In the exact polynomial model, the residue in Eq.~\eqref{hdPoleResidues} shows that, in the real-root domain, the additional pole is ghostlike under standard quantization; beyond that domain, the complex-conjugate poles do not represent ordinary asymptotic states. {In the truncated-EFT reading of Sec.~\ref{subsec:renormScheme} these poles lie outside the controlled domain and are not predictions.}}

\section{\revgreen{Conclusions}}
\label{sec:conclusions}

{We have analyzed a four-dimensional $\mathcal N=1$ supersymmetric Proca model defined by an explicit mass term for a real vector superfield and by its minimal gauge-covariant SQED-like coupling to chiral matter. In these Proca variables the theory has no gauge redundancy; for local kinetic operators it is the exact classical unitary-gauge form of the minimal Abelian $\mathcal N=1$ Stueckelberg realization, whose quantum-consistent gauge-invariant parent requires anomaly-free matter (Sec.~\ref{subsec:stueckelbergRealization}). {For the minimal model, the one-loop K\"ahlerian contribution to the effective action of chiral background superfields is obtained in agreement from the background-field Hessian and an independent supergraph organization; the analysis is then extended to a general transverse form factor.}

In the minimal model the background-dependent one-loop divergence is local and is absorbed by the matter K\"ahler normalization; dimensional reduction is used as the regulator and $\overline{\mathrm{DR}}$ as the subtraction scheme, and the finite local terms that the subtraction leaves are scheme dependent.} {At $\alpha=0$, and for an anomaly-free completion of the matter sector, the corresponding minimal Stueckelberg formulation supports a conditional perturbative-renormalizability interpretation. That reading combines the present off-shell K\"ahlerian result with the Stueckelberg formulation; it is not itself an on-shell renormalization statement, nor an all-order proof established here, and its conditions are set out in Sec.~\ref{subsec:renormScheme}.

For $f(\Box)=\Box+\alpha\Box^2$ the exact polynomial model, defined over the full momentum range, must be distinguished from a truncated derivative expansion valid only at low momenta, whose predictions do not extend to the ultraviolet-sensitive features of the exact model. The principal model-specific contribution is the one-loop K\"ahlerian determinant of an explicitly massive real vector superfield and its extension to a general transverse form factor, the ring-resummation technology itself being standard. The calculation is K\"ahlerian, so derivative-dependent and chiral contributions are not determined; it does not establish an all-order Ward identity, yields no conclusion about interacting spectral positivity, and establishes no renormalizability result for $\alpha\neq0$.

{\bf Acknowledgments.}
The work of A. C. L. was partially supported by CNPq under Grants No.~404310/2023-0 and No.~301256/2025-0. The work of A. Yu. P. was partially supported by CNPq under Grant No.~303777/2023-0.

\end{document}